\documentclass[manuscript,screen, nonacm]{acmart}
\AtBeginDocument{%
  \providecommand\BibTeX{{%
    \normalfont B\kern-0.5em{\scshape i\kern-0.25em b}\kern-0.8em\TeX}}}
\usepackage{float}
\usepackage{amsmath}
\usepackage{multirow}
\usepackage{array}
\usepackage{tabularx}
\usepackage{booktabs}
\usepackage{makecell}
\usepackage{xcolor}
\usepackage{graphicx}

\begin{document}


\title[Visual Data Obfuscation in Remote Proctoring]
{Balancing The Perception of Cheating Detection, Privacy and Fairness: A Mixed-Methods Study of Visual Data Obfuscation in Remote Proctoring}

\author{Suvadeep Mukherjee}
\email{suvadeep.mukherjee@uni.lu}
\orcid{0000-0002-1213-1767}
\affiliation{
  \institution{xCIT group, University of Luxembourg}
   \streetaddress{11 Porte des Science}
    \city{Esch-sur-Alzette}
  \country{Luxembourg}
}

\author{Verena Distler}
\email{verena.distler@unibw.de}
\orcid{0000-0002-4461-0551}
\affiliation{
  \institution{Universität der Bundeswehr München}
  \country{Germany}
}

\author{Gabriele Lenzini}
\email{gabriele.lenzini@uni.lu}
\orcid{0000-0001-8229-3270}
\affiliation{
    \institution{IRiSC group, University of Luxembourg}
  \country{Luxembourg}
}

\author{Pedro Cardoso-Leite}
\email{pedro.cardosoleite@uni.lu}
\orcid{0000-0002-2848-5527}
\affiliation{
  \institution{xCIT group, University of Luxembourg}
  \country{Luxembourg}
}








 \renewcommand{\shortauthors}{Mukherjee et al.}

\begin{abstract}

Remote proctoring technology, a cheating-preventive measure, often raises privacy and fairness concerns that may affect test-takers' experiences and the validity of test results.
Our study explores how selectively obfuscating information in video recordings can protect test-takers' privacy while ensuring effective and fair cheating detection. Interviews with experts (N=9) identified four key video regions indicative of potential cheating behaviors: the test-taker's face, body, background and the presence of individuals in the background.
Experts recommended specific obfuscation methods for each region based on privacy significance and cheating behavior frequency, ranging from conventional blurring to advanced methods like replacement with deepfake, 3D avatars and silhouetting.
We then conducted a vignette experiment with potential test-takers (N=259, non-experts) to evaluate their perceptions of cheating detection, visual privacy and fairness, using descriptions and examples of still images for each expert-recommended combination of video regions and obfuscation methods.
Our results indicate that the effectiveness of obfuscation methods varies by region. Tailoring remote proctoring with region-specific advanced obfuscation methods can improve the perceptions of privacy and fairness compared to the conventional methods, though it may decrease perceived information sufficiency for detecting cheating. However, non-experts preferred conventional blurring for videos they were more willing to share, highlighting a gap between the perceived effectiveness of the advanced obfuscation methods and their practical acceptance.
This study contributes to the field of user-centered privacy by suggesting promising directions to address current remote proctoring challenges and guiding future research.

\end{abstract}

\ccsdesc[500]{Human and societal aspects of security and privacy}
\ccsdesc[300]{Applied computing~E-learning}
\ccsdesc[300]{Applied computing~Distance learning}

\keywords{remote proctoring, cheating, obfuscation, privacy-utility, fairness, UX, willingness to share}


\maketitle

\section{Introduction}
In remote high-stakes testing, aimed at cost-effective competence assessment, cheating prevention is often done via video recording, known as remote proctoring. It comes in various forms, including human-involved (remote live invigilation or recorded sessions with post-test verification) and AI-driven (solely operated by algorithms) \cite{hussein2020evaluation}. Institutions often opt for a hybrid approach, combining algorithms to flag suspicious behaviors during tests with subsequent review of recordings by professionals \cite{li2021visual}, referred to as `reviewers' in this paper. Although remote proctoring is often mandated by institutions \cite{selwyn2023necessary} as anti-cheating measures, it can adversely affect test-takers' experiences due to significant privacy concerns \cite{balash2021examining, conijn2022fear, mukherjee2023effects}. This is because the video recordings may inadvertently disclose sensitive information about individuals (e.g., their family details, religious beliefs, disabilities or gender identities \cite{coghlan2021good, kharbat2021proctored}). Additionally, test-takers may feel uncertain about how their videos are observed by the reviewers \cite{terpstra2023online, coghlan2021good}. For instance, AI algorithms in proctoring software, often trained on biased datasets, might unjustly flag cheating incidents. When presented to reviewers, these biases can influence their judgments, potentially leading to unfair accusations of cheating \cite{burgess2022watching, yoder2022racial}.

This study aims to address these issues by focusing on concealing privacy-sensitive video content from reviewers to enhance test-takers' privacy perception. Leveraging recent advancements in computer vision, the study explores the scope of different obfuscation methods (i.e., video filters) used in various video conferencing tools (e.g., Zoom, Microsoft Teams) and social media platforms (e.g., Instagram, Snapchat) to alter appearances while preserving motion in videos. A range of methods, including blurring \cite{nousi2020deep}, pixelation \cite{hill2016effectiveness}, masking \cite{ravi2023review}, inpainting \cite{li2022obfuscation} and augmented reality 3D animations \cite{fu2008real}, can be used in live-streaming scenarios \cite{higgins2021remotely}. Additionally, more advanced methods like deepfake \cite{tolosana2020deepfakes}, which synthesizes realistic visual content, can be implemented using sophisticated video editing tools. This study also examines how obfuscation methods impact test-takers' perceptions of potential discrimination by reviewers, as they can conceal attributes like test-takers' ethnicity, age or gender \cite{roberts2014protecting}, potentially preventing unfair accusations \cite{yoder2022racial}. While these methods can protect privacy and reduce discrimination, they can also hide visual data crucial for identifying cheating behaviors like movements in head, mouth or eye gaze \cite{potluri2023automated}. This perceived reduction in cheating detection is crucial as it may influence test-takers' likelihood to cheat, regardless of the actual detection capabilities \cite{freiburger2017cheating}.

Hence, our research aims to: \emph{(1)} identify the most relevant obfuscation methods that can be applied in remote proctored videos, and \emph{(2)} investigate how potential test-takers perceive the effectiveness of those identified methods in terms of cheating information sufficiency, privacy protection and fairness in cheating detection. Additionally, 
to assess whether these methods improve user experience (UX), we measured the willingness of potential test-takers to share obfuscated videos with reviewers as an indicator of their trust and attitude toward these methods \cite{dwyer2007trust}. To the best of our knowledge, no prior research has investigated these three crucial aspects collectively when obfuscating video content in remote proctoring to enhance test-takers' experiences. For this purpose, we conducted a two-part study using a mixed-methods approach, combining a qualitative and quantitative study. 

The first part, addressing the following research questions, involved semi-structured interviews with experts to explore the significance of obfuscating specific regions (face, body, background and
other individuals) in a video.

\textbf{RQ1}: What specific visual information do experts believe different video regions offer for cheating detection?

\textbf{RQ2}: What obfuscation methods do experts believe should be relevant in each region to address test-takers' privacy concerns while ensuring sufficient information for cheating detection and avoiding unfair accusations?

\textbf{RQ3}: What recommendations do experts provide for applying the identified obfuscation methods on the video?

In the second part, to address the following research questions, we evaluated the effectiveness of the expert-recommended region-specific obfuscation methods with non-expert potential test-takers in a vignette experiment. To simplify the user-testing process, we used simulated images as a proxy for remote proctored videos with each of the obfuscation methods applied and asked participants to visualize them as visual scenes in the video recordings.

\textbf{RQ4}: What are the effects of the region-specific obfuscation methods on test-takers' perceptions of cheating information sufficiency, privacy protection and fairness in cheating detection?

\textbf{RQ5}: What are the effects of the region-specific obfuscation methods on these perceptions combined?

\textbf{RQ6}: What are the effects of the region-specific obfuscation methods on test-takers' willingness to share videos with unknown reviewers if obfuscated with those methods?

\subsection{Contributions}
\begin{itemize}
\item Our study provides a research direction toward improving test-taking experiences in remote proctoring by hiding sensitive visual data in recorded videos. The aim is to select obfuscation methods pragmatically not only for their capacity to balance perceptions of privacy protection and fair cheating detection but also for their potential to foster trust among test-takers, ensuring exam integrity while preserving privacy.
\item Our study proposes that obfuscation methods should be tailored to regions (face, body, background and other individuals) in a remote proctored video based on their privacy significance and frequency of cheating behavior.
\item We explored different obfuscation methods with distinct functions, eliciting different qualitative perceptions. Additionally, we highlighted fairness concerns to guide researchers considering their implementation in videos.
\end{itemize}

\label{sec:intro}

\section{Related Work}
\subsection{Cheating Detection in Remote Proctoring} \label{sec:cheating and proctoring}
In high-stakes remote testing, ensuring integrity is crucial for detecting cheating, such as unauthorized resource use or external assistance. Remote proctoring, a common cheating-preventive method, typically involves three setups, each with its own challenges \cite{hussein2020evaluation}. The first involves live monitoring by remote invigilators, who intervene
immediately for exam rule violations. This is resource-intensive and lacks test-taking flexibility. The second records the entire test session for later professional review to identify cheating behaviors by watching the entire video, but it's time-consuming. The third relies on AI/ML algorithms alerting test-takers if a suspicious behavior is detected during test-taking, but raises concerns due to the nature of
the datasets the algorithms are trained with. Test-organizing institutes often use a hybrid approach \cite{li2021visual} to address these shortcomings, ensuring cost-effectiveness. The process includes recording the test-taking session, with algorithms generating reports of detected cheating behavior \cite{potluri2023automated} and sending these reports to reviewers afterward, enabling them to access relevant timestamps, hence streamlining the process \cite{terpstra2023online}. Various suspicious events (e.g., test-takers being absent from the frame, the presence of other individuals, different individuals taking the test and students disabling the webcam) have been documented as potential cheating behaviors \cite{yoder2022racial}. Our study focuses on the hybrid setup, promising for obfuscation applications and streamlining cheating verification.

\subsection{Privacy Concerns with Video Recordings in Remote Proctoring} \label{sec:privacy}
In recent times, concerns over privacy perception have surged due to the abundance of detailed visual content captured by video recording devices. From public surveillance cameras capturing individuals without consent \cite{burt2019, whitelaw2020applications} to the use of remote proctoring in high-stakes testing, mandated by institutional obligations \cite{selwyn2023necessary}, unintended recording of sensitive information like test-takers' family members \cite{kagan2023zooming}, religious affiliations \cite{kharbat2021proctored}, disabilities, or gender orientations \cite{coghlan2021good} may not be considered necessary for ensuring test integrity. Despite encryption and access control measures, increasing online data breaches have heightened user concerns about personal data privacy \cite{dadashzadeh2021online, james2017exposing}, potentially leading to resistance \cite{selwyn2023necessary} and legal actions \cite{witley2023, meaker2023}. Therefore, privacy-enhancing measures should be perceived as robust strategies by testing stakeholders, particularly by test-takers, as they can influence test-taking experiences \cite{balash2021examining} and subsequent adoption of the practice \cite{ul2023nuanced}. These measures should reassure test-takers that their videos are not inappropriately monitored or pose harm if shared with third parties \cite{coghlan2021good}, preventing unnecessary privacy concerns and distress \cite{balash2021examining}.

\subsection{Fairness Concerns with Cheating Detection Outcome in Remote Proctoring} \label{sec:fairness}
In hybrid remote proctoring, where reviewers consult algorithm-generated reports of detected suspicious behaviors before verifying corresponding video timestamps, risks might arise in terms of fair cheating detection. For instance, if reviewers prioritize algorithm reliance over objective judgment, unfair cheating accusations can occur due to algorithmic biases from skewed training datasets \cite{yoder2022racial, burgess2022watching}. Studies \cite{yoder2022racial} indicate that algorithms tend to flag individuals with darker skin tones more frequently for cheating allegations, particularly affecting females with dark skin tones. Moreover, algorithms may flag unconventional student actions during tests (e.g., unusual head movements, muttering, looking to the side, leaning on hands, wiping faces, drinking water or approaching the screen closely) leading to increased false positives \cite{yoder2022racial, li2021visual}. Additionally, reviewers' personal biases, influenced by factors like test-takers' ethnicity, skin tone and gender, can also result in unfair accusations \cite{grindstaff2019no}. Given these limitations, a privacy-preserving measure should provide assurance to test-takers that such biases will not impact the fairness of cheating detection outcomes.

\subsection{Obfuscating Video Contents in Remote Proctoring}
Privacy protection in video-based applications often involves obscuring identifiable video content. Recent advancements in AI and computer vision allow for real-time alteration of appearances in video chats or post-editing of recorded videos using visual obfuscation methods like facial filters and dynamic augmented reality animations, widely popular on social media (e.g., Snapchat, Instagram etc.) and increasingly integrated into video conferencing platforms (e.g., Zoom, Microsoft Teams). Some of the prevalent methods include conventional and lightweight methods like \textit{blurring} \cite{nousi2020deep} and \textit{pixelation} \cite{hill2016effectiveness} that alter a region by recalculating neighboring pixel values; \textit{masking} with a solid box \cite{ravi2023review}; \textit{inpainting} to fill missing parts \cite{li2022obfuscation}; to more advanced methods like \textit{deepfake} technology for synthesizing video content \cite{tolosana2020deepfakes} or replacing individuals with \textit{3D cartoon avatars} \cite{fu2008real} and \textit{silhouette-like} figures by contour masking \cite{li2017effectiveness}. While concealing identifiable video content can improve the user experience by preserving privacy perception \cite{javornik2022lies}, the selection of methods often relies on the objective evaluations of various factors within specific contexts \cite{erdelyi2018privacy, ravi2023review}. 

One key consideration is the region of the video where obfuscation is applied; for example, studies show that hiding the entire body enhances privacy perception more than just hiding the face \cite{chen2009protecting}, and masking is more effective than blurring in that task \cite{li2017effectiveness}. However, it can also lead to the loss of crucial details relevant to the context; for instance, masking of test-takers in remote proctored videos can obscure vital behavioral cues like body movement or eye gaze, underscoring the importance of balancing privacy and utility \cite{ravi2023review}. Obfuscation methods also vary in their computational demands \cite{zhuang2018performance}; for instance, applying deepfake to synthesize the test-taker's features during video editing could demand high processing power, affecting test institutions' budgets. Furthermore, evaluating methods often involves assessing their effectiveness against not only human observation (perceptual obfuscation) but also adversarial algorithmic attacks aimed at reversing obfuscation (machine obfuscation) \cite{ravi2023review}, such as the potential identity reversibility seen in blurring and pixelation \cite{kupyn2019deblurgan}. Given the various factors influencing such evaluations, there is a need to pragmatically select obfuscation methods, which is explored in Section~\ref{sec:study1} followed by user evaluation in Section~\ref{sec:study2}.

\subsection{User Experience in Remote Proctoring} 
In recent years, assessing user experience (UX) has gained popularity to measure technology's impact on users by going beyond usability (i.e., ease of use, efficiency) and including emotional responses, trust and beliefs resulting from user interaction with a digital product \cite{hassenzahl2006user}. Technologies requiring users to share personal information often face adoption challenges \cite{ul2023nuanced} due to concerns about transparency in data usage and potential misuse \cite{kim2017meta}, as well as the risk of discrimination based on identifiable data, impacting user trusts \cite{yang2022user}. In remote proctoring, where test-takers share test-session videos under obligatory conditions amid resistance due to poor test-taking experiences \cite{selwyn2023necessary, coghlan2021good}, assessing their trust in obfuscation methods applied to video content is crucial for UX, alongside their objective evaluations for privacy safeguards and fair cheating detection. Measuring trust through their willingness to share video \cite{ul2023nuanced} under applied obfuscation can guide us assessing adoption and hence standardizing video recording for test integrity.
\label{sec:sota}

\section{Study Design}

This study, divided into two parts, aims to identify the most promising obfuscation methods for remote proctoring video recordings and to determine how these methods, by hiding sensitive and potentially discriminatory data in the videos, affect test-takers' perception of privacy protection and fairness in cheating detection. Simultaneously, the obfuscation methods must not compromise cheating information sufficiency, as it can influence their test-taking experience and likelihood of cheating, assuming they believe that cheating actions can be identified. In Part 1, experts recommended obfuscation methods for hiding specific video regions, known as Regions of Interest (ROIs). Part 2 evaluated these recommended methods with potential test-takers, assessing their perceptions of privacy, fairness and information sufficiency, as well as their willingness to adopt each method for remote proctoring.\label{sec:RO}

\section{Part 1: Identifying Suitable Obfuscation Methods in Remote Proctoring}\label{sec:study1}

This section gathers insights from expert interviews regarding factors to consider when identifying obfuscation methods relevant for hiding specific video regions. It also aims to propose promising remote proctoring pipeline, such as if obfuscation should occur in real time or post-test and the associated tasks of different test stakeholders.

\subsection{Recruitment, Interview Protocol and Analysis}
For interview purpose, we recruited nine experts based in the USA and Europe through email within the researchers' network (details are provided in Appendix Table~\ref{tab:experts-profile}). This group comprised three professionals with expertise in remote proctoring within universities, four specialists engaged in computer vision research in the industry and academia and two researchers specialized in usable privacy, also serving as professors in universities. Each expert engaged in a session lasting 1-1.5 hours and received €40 for their time. The sessions were conducted remotely in a semi-structured format.

The first author facilitated expert engagement via an online collaboration platform, `miro.com'. Interview materials are provided in Appendix Figure~\ref{fig:task}. Initially, experts identified visual cues indicating potential cheating behaviors in different regions (faces, bodies, backgrounds) of a visual scene by examining a simulated image of front-facing test-takers with visible body parts. They rated the importance of these cues on a scale from 1 (least) to 10 (most) for cheating detection. Next, they evaluated five pre-prepared images with various obfuscation methods applied to faces (blurred, pixelated, masked, deepfaked avatarized), placing them on a 2D-MAP with the x-axis representing privacy protection and the y-axis as cheating detection difficulty, while elaborating on their decisions. This process was repeated by asking experts to visualize similar obfuscation methods applied to other regions. Experts also suggested measures to mitigate potential biases by hiding potentially discriminatory attributes to ensure fair judgments in cheating detection. Finally, experts proposed a viable solution for a cost-effective obfuscation pipeline for remote proctoring.

To extract insights, the first author analyzed both task outcomes on `miro.com' and related discussions. The interviews were transcribed and analyzed using deductive content analysis in MAXQDA (v.2024). Initial coding involved five main categories: identifying cheating behaviors for each region, assessing the importance of visual cues for detecting cheating behaviors, evaluating the advantages and disadvantages of obfuscation methods in each region, proposing measures to address fairness concerns and discussing the obfuscation pipeline. The transcripts were thoroughly reviewed and the relevant content were highlighted, coded and categorized accordingly. The codebook is provided in Appendix~\ref{appdx:qualcodes}.

\subsection{Results} \label{expert-result}
\subsubsection{Significance of Regions of Interest (ROI) for Cheating Detection in Visual Scenes}

The focus was on distinguishing possible cheating instances linked to foreground (facial and body regions) and background (stationary and moving elements like individuals appearing behind during tests) regions or ROIs.

Most experts, particularly those with experience in remote proctoring, identified that the most common and frequent cheating activities originate from the face region, with indicators such as mouth, head and eye movement being cited as the most frequent. This underscores the reviewers' need for significant attention in this area. Almost all experts stressed that potential cheating instances within an ROI shouldn't be viewed in isolation, as cheating behavior is multifaceted and may involve various cues. For example, body movement accompanied by a change in eye gaze direction or a synchronized mouth movement between a test-taker and a person in the background might indicate suspicious behavior. All identified cheating instances to each ROI, along with potential associations with other ROIs, are presented in Table~\ref{tab:cheating_ROI}. Beyond test-takers' face and body, the background region may also serve to detect cheating, for example, revealing unpermitted resources or interactions with other people. Remote testing guidelines often advise on suitable test-taking locations to prevent such disclosures, though enforcement may vary given the diverse living situations of test-takers.

\begin{table}[h]
  \scriptsize
  \caption{List of potential cheating occurrences linked to regions of interest (ROI)}
  \label{tab:cheating_ROI}
  \begin{tabular}{llll}
    \hline
    \textbf{\makecell[c]{Concerned \\ROI}} & \textbf{\makecell[c]{Cues for \\suspicious\\ behavior}} & \makecell[c]{\textbf{Cheating instances}} & \textbf{\makecell[c]{Possible association \\to other ROIs}} \\
    \hline
    \multirow{4}{*}{Face} & Mouth movement & Discussing answers using an earphone; asking for answers from someone present & \multirow{4}{*}{\makecell[l]{People in \\background}}\\
    \cmidrule(lr){2-3}
    & Eye movement & Looking away from the screen; interacting with someone in the room &  \\
    \cmidrule(lr){2-3}
    & Head movement & Interacting with someone in the room; lowering head to the desk, possibly using unauthorized materials & \\
    \hline
    \multirow{4}{*}{Body} & Body pose & Moving away from the screen; interacting with someone present & \multirow{4}{*}{\makecell[l]{People in \\background}} \\
    \cmidrule(lr){2-3}
    & Hand movement & Using unauthorized materials (e.g., smartphone); interacting with objects like books & \\
    \cmidrule(lr){2-3}
    & Shoulder movement & Shifting shoulders to engage in cheating activities such as using phones, books, etc. & \\ 
    \hline
    \multirow{3}{*}{Background} & Presence of camera & Live-feeding computer screen for remote question dissemination & \\
    \cmidrule(lr){2-3}
    & Visible cheat notes & Test-taker writes cryptic answers on posters or walls & \\
    \hline
    \multirow{3}{*}{\makecell[l]{People in\\ background}} & Mouth movement & Talking to the test-taker during the exam & \multirow{3}{*}{Face, body}  \\
    \cmidrule(lr){2-3}
    & Body movement & Approaching test-takers closely to assist or view the computer screen & \\
    \hline
  \end{tabular}
\end{table}

\subsubsection{Obfuscation Methods to Balance Privacy Protection and Cheating Detection}

Following the identification of potential cheating instances for each ROI, experts evaluated common obfuscation methods (e.g., blurring, pixelation, masking, 3D avatar representation and deepfake), focusing primarily on privacy significance and the frequency of cheating behavior in each ROI. Additionally, factors such as their effectiveness in improving privacy protection, retaining cheating information post-obfuscation, scope of reversibility by de-obfuscating algorithms and practical challenges like computational demands when preserving motion for faces, bodies etc. \cite{ravi2023review}, were also taken into account.

Experts observed that once test-takers are authenticated before the test, facial and body features become less crucial for review, allowing for their obfuscation during video editing, provided that motion from these areas, such as eye and mouth movement or hand gestures, is adequately preserved. Conventional obfuscation like blurring and pixelation were considered moderately effective for privacy protection and cheating information preservation across all ROIs, but masking was deemed unsuitable for the face and body regions due to eliminating crucial indicators of suspicious behavior and inability to preserve motion. For the face region, deepfake and 3D avatar replacement were preferred over other methods due to their superior privacy protection (in terms of both perceptual and identity irreversibility \cite{yuan2022pro, ravi2023review, tolosana2020deepfakes, kupyn2019deblurgan}) and ability to retain most cheating information if motion is preserved. However, computer vision experts warned that they can be computationally more intensive than conventional obfuscation methods, raising concerns over institutional support and budget for video editing. Deepfaked face was somewhat preferred over 3D avatar replacement because the former offers more realistic identity replacement by providing more texture information \cite{wang2020learning}. For the body region, a straightforward approach like outfit replacement can be effective, especially if motion, including hand and shoulder movement and changes in body poses, is well preserved. For that purpose, blurring, pixelation and deepfake were rated by experts as viable options. 
To obfuscate infrequent instances like people appearing in the background, a variant of masking, such as silhouette-like figure replacement based on region contour, was suggested, as it captures suspicious body behavior while overlooking individual indicators like eye and mouth movements. Alternatively, a motion-preserving full-body 3D avatar could also be effective. To obfuscate the background, blurring, pixelation or replacing it with a picture can achieve the goal of concealing stationary objects. Table~\ref{tab:obfuscation-methods} summarizes the expert recommendations for obfuscation methods in each ROI. Note that, we opted for blurring over pixelation for our user evaluation phase in Section~\ref{sec:study2} because of their similar effectiveness \cite{lander2001evaluating}, thus avoiding redundancy in user assessment.


\subsubsection{Addressing Potential Fairness Concerns in Cheating Detection}
Obfuscating face, body and other ROIs to enhance privacy protection may simultaneously suppress discriminatory attributes, potentially mitigating unfair judgments influenced by reviewers' personal biases. However, concerns about fairness may still arise regarding other discriminatory cues such as skin tone, ethnicity and gender of test-takers \cite{grindstaff2019no}. Privacy experts raised concerns about the possible interdependence between ethnicity and skin tone \cite{yoder2022racial}, suggesting that altering one without the other might not ensure fairness. Mixed opinions emerged regarding changing the skin tone when replacing a face with a different identity, prompting user-testing with altered skin tones. Caution was also advised to ensure consistency by extending alterations to other visible skin areas (e.g., neck, hands). Moreover, replacing all faces with a single ethnicity though may reduce discrimination, caution is needed regarding factors like ethnic attire or ornaments that could still reveal ethnicity. Addressing gender bias, including cues like hair textures and length, also requires careful consideration.

\begin{table}[H]
    \begin{minipage}{0.44\textwidth}
        \centering
        \scriptsize
        \begin{tabular}{lcccc}
        \hline
        \textbf{\makecell[l]{Region of \\interest}}& \textbf{Blurring}& \textbf{Silhouette}
        &\textbf{Deepfake}
        &\textbf{3D avatar}\\ \hline
    \makecell[l]{Face} & \checkmark & X & \checkmark & \checkmark\\
    \makecell[l]{Body}  & \checkmark & X & \checkmark & X\\
    \makecell[l]{Background}  & \checkmark & X & \checkmark & X\\
    \makecell[l]{People in \\background}  & \checkmark & \checkmark & X & \checkmark\\ \hline
        \end{tabular}
        \captionof{table}{Relevant region-specific obfuscation methods as recommended by experts}
        \label{tab:obfuscation-methods}
    \end{minipage}
    \hfill
    \begin{minipage}{0.48\textwidth}
        \centering
        \includegraphics[width=\textwidth]{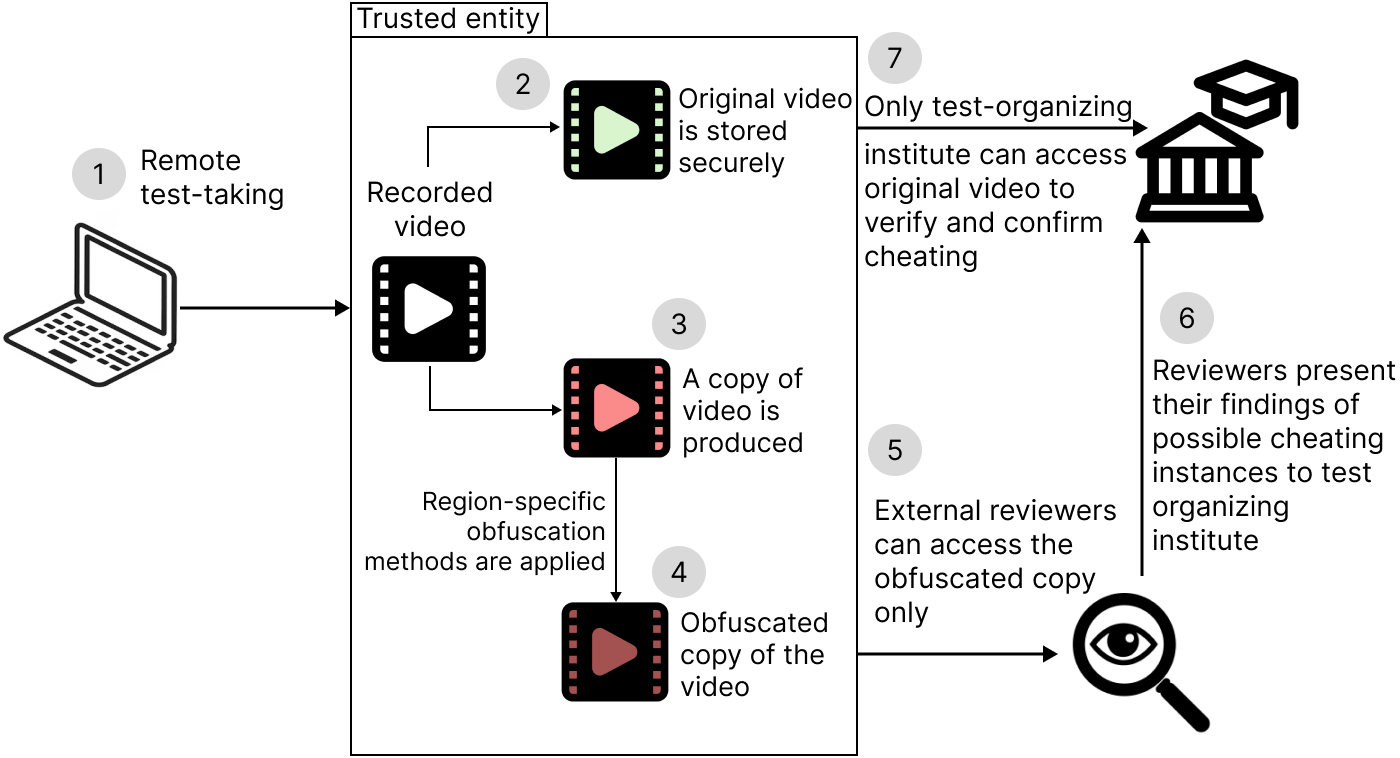}
        \captionof{figure}{Expert recommended pipeline for video obfuscation}
        \label{fig:pipeline}
    \end{minipage}
\end{table}

\subsubsection{Practical Obfuscation Pipeline for Remote Proctored Videos} \label{sec:pipeline}
Considering the practical limitations of obfuscation, nearly all experts stressed that relying solely on obfuscated videos for detecting cheating could pose significant challenges during disputes over cheating allegations. To tackle this, they suggested securely storing an unprocessed copy of the recorded video by a trusted entity. Region-specific obfuscation should then be applied to the original video there, with access granted for review purposes. The unprocessed video should be restricted to the test-organizing institutions, enabling them to make final decisions \cite{yaqub2022privacy}. The visual representation of the pipeline is depicted in Figure~\ref{fig:pipeline}.

\section{Part 2: Non-expert User Evaluation of Obfuscation Methods in Remote Proctoring}\label{sec:study2}


\subsection{Methodology}

\subsubsection{Experimental Design}
\begin{figure}[h]
  \centering
 \includegraphics[width=0.95\linewidth]{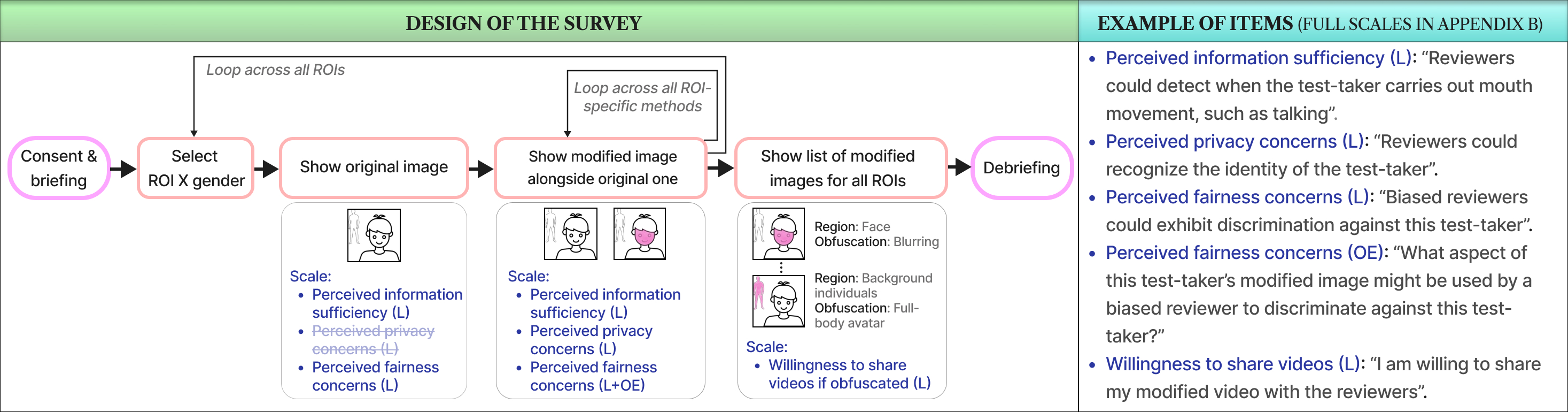}
  \caption{The survey presents an image as a visual scene, followed by hiding a region using various obfuscation methods. For instance, the test-taker's face above is highlighted to indicate where obfuscation is applied. Participants then provided their opinions using Likert scale (L) or open-ended (OE) items, with example items shown on the right side. Genders varied over the ROI in focus: for face, body and background obfuscation, test-takers' genders were varied; for background people, the genders of those people were varied}
\label{fig:studydesign}
\end{figure}

We conducted a vignette experiment with 259 non-expert potential test-takers to evaluate expert-recommended obfuscation methods on perceptions of privacy protection, fairness, cheating information sufficiency and user experience (UX). We used a 4x4 within-subject design with four obfuscation methods (blurring, silhouette, deepfake and 3D avatar) across four ROIs: face, body, background and people in the background. However, we only tested obfuscation methods that are specific to an ROI as recommended by experts (see Table~\ref{tab:obfuscation-methods}). Participants first received a brief introduction to the current challenges of video recordings followed by our research objectives, an example stimulus demonstrating obfuscation, and task instructions. The experimental design is depicted in Figure~\ref{fig:studydesign}. Each ROI was presented with two gender variations, showing original stimuli followed by region-specific obfuscations alongside the originals, and participants completed various questionnaires (details are in Section~\ref{sec:questionnaire}) accordingly.

In the survey, we used still images as test stimuli instead of video for simplicity. Participants were instructed to visualize the image as a visual scene in a video recording. This approach was chosen for several reasons. Firstly, research \cite{roberts2009static, conway2008evidence} indicates that static images or frames, which are sequenced to create videos, can capture essential aspects influencing user evaluations, even though they may not fully represent the complexities and nuances of dynamic video content. 
Additionally, static images might allow for better systematic control over variables in a visual scene compared to video-based stimuli \cite{penton2008attractiveness}, where a subtle change in context and how participants watch them (skipping or watching in entirety)  can introduce variability in judgments \cite{yang2022science}.

\subsubsection{Creation of Stimuli}\label{sec:stimuli}
\begin{figure*}[h]
  \centering
\includegraphics[width=0.95\linewidth]{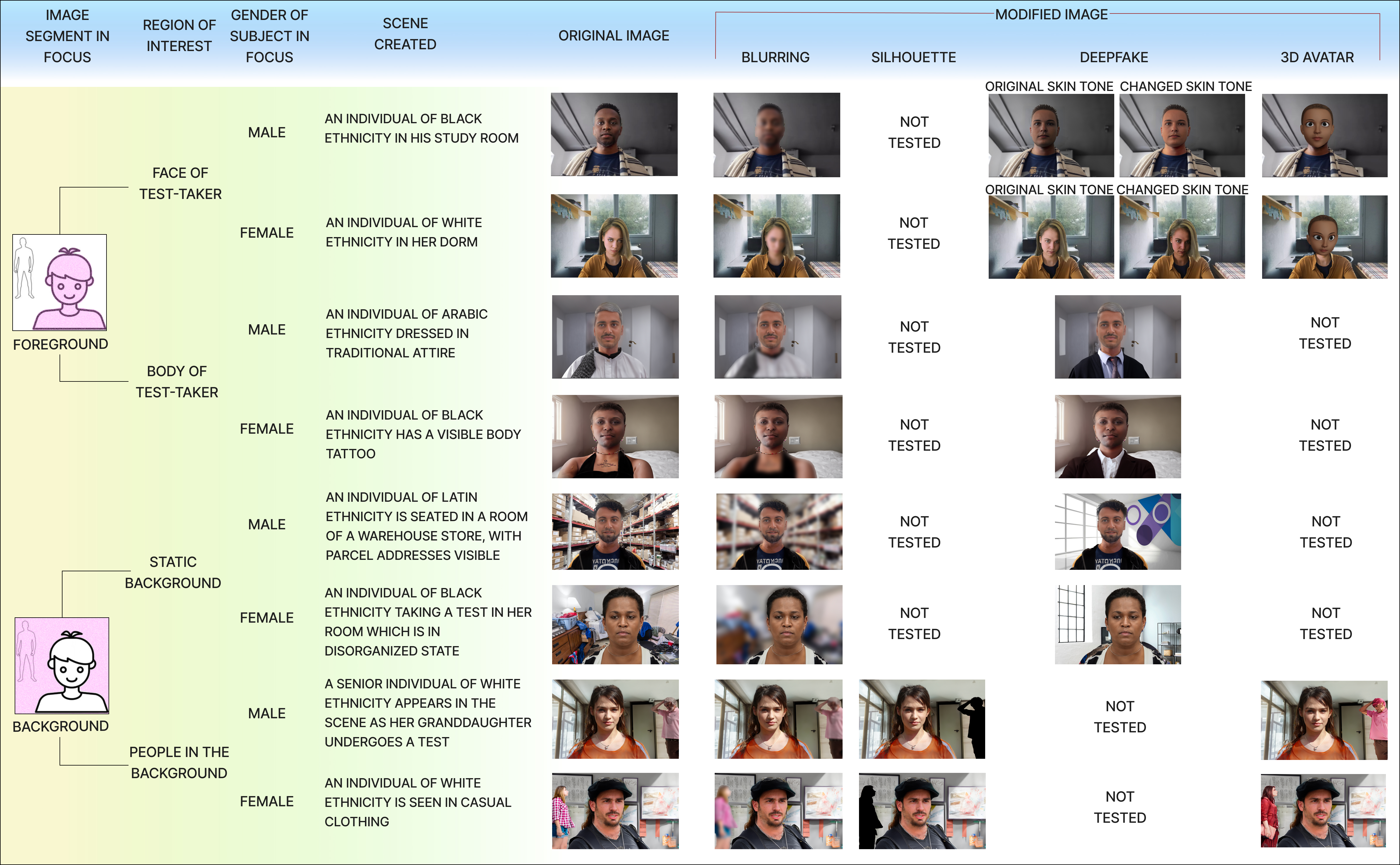}
  \caption{Visual scenes are created as stimuli for our vignette experiment, manipulating four regions (face, body, background people in the background) with relevant obfuscation methods. Both male and female subjects are represented varied across regions in focus, resulting in a total of 8 visual scenes. When deepfake was applied to the facial region, the questionnaire was assessed for both the original and changed skin tones of the subjects}
  \label{fig:stimuli}
\end{figure*}

The objective was to create set of original images and apply obfuscation methods to them.

\textbf{Creation of original images:} The first step involved generating original images of test-takers in simulated webcam-captured scenarios, ensuring equal gender representation, featuring diverse skin tones and diverse test-taking environments such as personal rooms or office spaces. Faces, aged between 18-40 years, sourced from \emph{thispersondoesnotexist.com}, were chosen for their realistic yet non-existent appearance, while Adobe Photoshop's (ver. 2024) Generative Fill feature was used to create other elements like the body and background. Faces from various ethnicities, including black, white, Arabic and Latin were included, resulting in two gender variations for each of the four ROIs, totaling 8 original images (refer to Figure~\ref{fig:stimuli}), with ethnicities randomly distributed. 


\textbf{Creation of modified images:} 
The second step involved creating an augmented set of stimuli by obfuscating the original images, following expert recommendations for region-specific obfuscation (see Table~\ref{tab:obfuscation-methods}). This process, varied over two genders, resulted in 20 modified images. Additionally, two variations of deepfaked faces were created to evaluate potential fairness concerns regarding skin tone, resulting in a total of 22 images (refer to Figure~\ref{fig:stimuli}). Next, we detail the various obfuscation methods employed.

\emph{1. \textbf{Blurring}}: We applied Gaussian blur with varying radii to balance cheating information sufficiency and privacy. Radii of 14 pixels for the face, 50 pixels for the body, 20 pixels for the background and 40 pixels for people in the background were found to strike this balance in pilot testing (see Appendix~\ref{appdx:pilot}).
\emph{2. \textbf{Deepfake}}: To address potential gender bias in facial obfuscation, we opted for a gender-neutral reference face created with the FaceMaker application \cite{schwind2015determining}. Using open-source deepfake tools \cite{roop}, we swapped faces while maintaining original facial expressions' fidelity. To evaluate fairness \cite{thong2023beyond}, each deepfaked face had two versions: one preserved the original skin tone, while the other was altered to the category 3 in the Fitzpatrick scale \cite{fitzpatrick1988validity}, a widely accepted skin color standard. Additionally, we dressed test-takers uniformly in formal attire, adjusted slightly for each gender, and replaced backgrounds with freely available room pictures. 
\emph{3. \textbf{3D avatar}}: In contrast to realistic deepfake identities, cartoonized avatars provide less detailed visual information \cite{wang2020learning}. Using the FaceMaker application \cite{schwind2015determining}, we generated gender-neutral cartoonized 3D face avatars. These avatars replaced the original faces, with attempts made to align them with the facial pose and expressions using Photoshop. Similarly, for individuals in the background, the visible portions were altered to 3D full-body avatars, mimicking the subjects' facial and body expressions, using the Generative Fill feature of Photoshop.
\emph{4. \textbf{Silhouette}}: We used a 2D silhouette representation only for the individuals in the background, filled with black color.

\subsubsection{Measurements} \label{sec:questionnaire}
Participants were shown 30 images: 8 original and 22 modified. For each modified image shown alongside its original (as depicted in Figure~\ref{fig:studydesign}), participants rated items on the perception of cheating information sufficiency, privacy concerns and fairness concerns using 7-point Likert scales from "strongly disagree" to "strongly agree". Fairness concerns also included an open-ended question to gauge potential bias post-obfuscation. All items were prepared based on expert interview findings (details are in Appendix Table~\ref{tab:measurements}). The 8 original images, without any obfuscation, served as the reference for comparing the effectiveness of obfuscation methods in later sections. For those images, privacy concern items were skipped because they were tailored to assess reviewers' ability to recognize specific ROIs, which we assumed they could easily identify without obfuscation. In contrast, items for information sufficiency and fairness concerns were determined through participant responses, reflecting their subjective nature, even without obfuscation.  After assessing the 22 images, participants rated their willingness to share videos with a reviewer for each obfuscation method using a similar Likert scale, measuring UX and preferences \cite{li2017effectiveness, li2022obfuscation} for sharing visual data.

\subsubsection{Recruitment and Ethical Considerations}
Since the study aimed to evaluate the broad impressions of region-specific obfuscation methods to achieve a balance between privacy protection and fair cheating detection, we conducted user testing with 259 UK-based remote participants recruited via Prolific, a reliable crowd-working platform known for providing high-quality data \cite{peer2022data}. This sample size ensured robust statistical inference, crucial for guiding potential applications in remote proctored videos. Adult participants aged 18 to 60 were considered eligible and were then directed to the survey. Data collection took place in January 2024. On average, participants took 25 minutes to complete the survey. The sample was non-representative, with 51.3\% female, 46.7\% male, and 2\% non-binary participants. On average, participants were 32 years old (SD=10) with diverse educational backgrounds and around 67\% had a university degree. Approximately 37\% had taken at least one remote proctored test in the last three years.

At the beginning of the survey, participants were given a digital informed consent form and a study information sheet. We followed GDPR practices and informed participants about data collection, storage and opt-out opportunities. Following completion, participants received a written debriefing explaining the study's purpose and the creation of images for research purposes. Each participant received compensation as per Prolific's hourly pay policy. The university’s ethics committee reviewed and approved our research project.

\begin{table}[h]
\scriptsize
    \centering
    \caption{Mean values with std. deviations of all dependent variables for each combination of ROIs and obfuscation methods. The means of perceived privacy protection in `No obfuscation' were considered 1.00 for analysis purposes, as explained in Section~\ref{sec:analysis}}
    \label{tab:descriptive}
    \begin{tabular}{clccccc}
        \toprule
        {\makecell{\textbf{Region of} \\ \textbf{interest}}} & {\makecell[l]{\textbf{Obfuscation}\\\textbf{methods}}} & {\makecell{\textbf{Perceived informa-}\\\textbf{tion sufficiency}}}  & {\makecell{\textbf{Perceived privacy}\\ \textbf{protection}}}  &  {\makecell{\textbf{Perceived}\\ \textbf{fairness}}}  & {\makecell{\textbf{Composite} \\\textbf{scores}}} & {\makecell{\textbf{Willingness to} \\\textbf{share videos}}}\\
        \midrule
        
        \multirow{5}{*}{\makecell{Face of \\ test-taker}} & No obfuscation & 6.14 (0.85) & -  & 2.25 (1.13)  & 9.39 (1.17) & -\\
        
        & Blurring  & 4.57 (1.32) & 4.21 (1.45)  & 2.57 (1.14)  & 11.34 (2.01) & 4.94 (1.71)\\
        
        & Deepfake (Original skin tone) & 5.68 (1.02) & 4.67 1.38) & 3.12 (1.22)  & 13.46 (1.84) & 3.42 (1.87)\\

        & Deepfake (Changed skin tone)  & 5.67 (1.04) & 4.66 (1.38)& 3.16 (1.15)  & 13.51 (1.84) & 2.78 (1.61)\\

        & 3D avatar  & 4.65 (1.51) & 6.16 (1.04)& 3.94 (1.52)  & 14.76 (2.23) & 4.13 (2.13)\\

        \midrule

        \multirow{3}{*}
        {\makecell{Body of \\ test-taker}} & No obfuscation   & 5.56 (1.04) & - & 2.46 (1.08)  & 9.01 (1.35) & -\\    

        & Blurring  & 4.99 (1.13) & 4.27 (1.21)  & 3.61 (1.33)  & 12.87 (1.99) & 4.49 (1.89)\\
        
        & Deepfake & 5.22 (1.08) & 5.09 (1.51) & 4.31 (1.72)  & 14.63 (2.63) & 3.97 (2.02)\\
        
        \midrule
        
        \multirow{3}{*}
        {\makecell{Background \\ of test-taker}} & No obfuscation   & 5.23 (1.19) & - & 3.61 (0.66)  & 9.83 (1.43) & -\\

        & Blurring  & 4.21 (1.25) & 3.59 (1.28)  & 3.51 (1.39)  & 11.29 (2.17) & 5.12 (1.73)\\
        
        & Deepfake & 3.56 (1.55) & 6.06 (1.33) & 5.45 (1.55)  & 15.07 (2.24) & 5.23 (1.78)\\
         
        \midrule
        
        \multirow{4}{*}{\makecell{People in \\ background}} & No obfuscation  & 5.96 (0.94) & - & 2.69 (1.42) & 9.65 (1.59) & -\\
        
        & Blurring  & 4.58 (1.37) & 4.62 (1.64)  & 3.41 (1.55)  & 12.62 (2.42) & 4.88 (1.91)\\
        
        & Silhouette   & 3.85 (1.47) & 5.97 (1.42) & 4.49 (1.87)  & 14.31 (2.45) & 4.49 (2.02)\\

        & 3D avatar   & 4.95 (1.28) & 5.07 (1.63) & 3.67 (1.65)  & 13.69 (2.52) & 3.62 (2.03)\\
 
    \bottomrule
    \end{tabular}
\end{table}

\subsubsection{Data Analysis} \label{sec:analysis} 
In the following section, we address three research questions: \emph{(1) RQ4} investigates the effect of region-specific obfuscation methods on perceptions of information sufficiency, privacy and fairness; \emph{(2) RQ5} examines their effect on the overall perceptions; and \emph{(3) RQ6} investigates their effect on participants' willingness to share videos with unknown reviewers if obfuscated. Following data collection, we prepared the dataset for dependent variables in three steps. First, we assigned a maximum value of 7 uniformly to all participants for the skipped privacy concerns items as explained in Section~\ref{sec:questionnaire}. We then reversed the Likert scale responses for privacy and fairness concerns to align with our research objectives, considering a rating of 1 as 7 and vice versa, with higher scores indicating greater privacy protection and fairness. Second, we computed the mean values of the dependent variables for each image if multiple items were asked within each ROI. Finally, we averaged the ratings across the two image variations shown per ROI. Table~\ref{tab:descriptive} presents the data prepared for the dependent variables used for statistical analysis and visual representations.

To statistically compare the effect of obfuscation methods on the dependent variables for each ROI, we conducted linear mixed-effects models (LMEM) followed by post hoc analyses using Wald tests. The independent variables for the models were the types of obfuscation methods per ROI, with the baseline being the ratings for `no obfuscation' condition. The mixed-effects models were considered to account for variability in participants' responses, considering multiple ratings for the methods per participant. All statistical analyses were conducted using STATA v18.  The covariance structure in LMEM models was set as `unstructured' to capture correlations among random effects of participants.

\begin{figure}[h]
  \centering
  \includegraphics[width=0.8\linewidth]{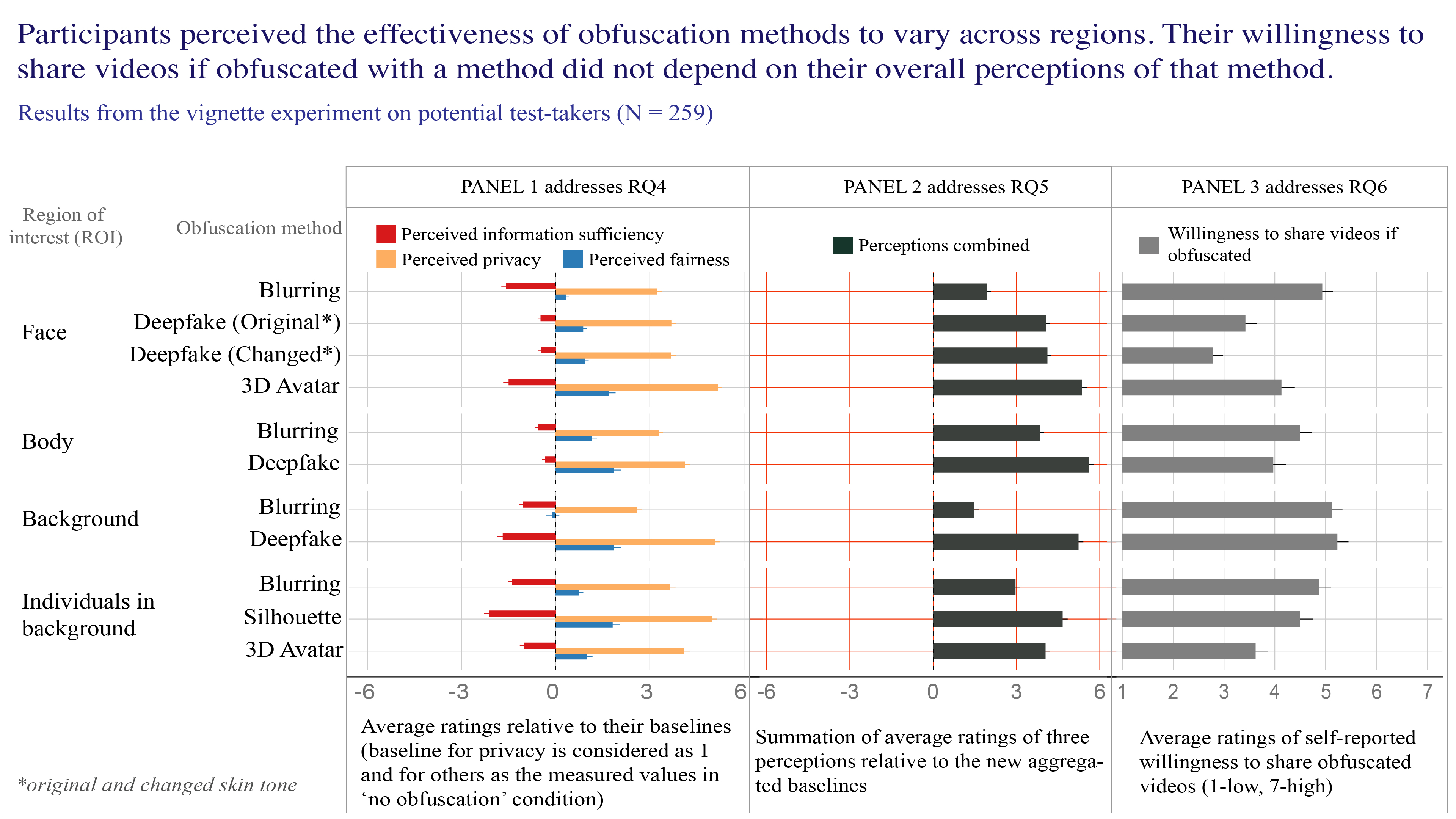}
  \caption{Three panels address three research questions respectively: the impact of region-specific obfuscation methods on \emph{(RQ4)} perceptions of information sufficiency, privacy and fairness; \emph{(RQ5)} the combined perception; and \emph{(RQ6)} willingness to share videos if obfuscated. PANEL 1 and PANEL 2 plot bars relative to the respective baseline values}
\label{fig:allthree}
\end{figure}

\subsection{Results} \label{result}

\subsubsection{Effects of Region-specific Obfuscation Methods on Perceived Information Sufficiency, Perceived Privacy and Perceived Fairness}\label{sec:individual_effects}

Based on the prepared data in Table~\ref{tab:descriptive}, PANEL 1 in Figure~\ref{fig:allthree} illustrates the average ratings for perceptions of information sufficiency, privacy and fairness relative to `no obfuscation' condition for each combination of ROIs and obfuscation methods. These perceptions varied across both ROIs and methods. For instance, replacing a test-taker's face with a 3D avatar might be perceived as better for perceived privacy and fairness than using a realistic deepfake face, but it could compromise more cheating information. To statistically validate the findings from the figure, we conducted 12 mixed-effects models for three dependent variables across four ROIs, followed by post hoc pairwise comparisons. The detailed results can be seen in Appendix Table~\ref{tab:statistics} and \ref{tab:waldresult}. Overall, the variability in dependent variables in the models was relatively low \emph{(<1.42)} across all ROIs, suggesting that the primary source of variability was the impact of obfuscation methods rather than individual differences. The discussion below presents some key findings for the foreground (test-takers' face and body) and background (background and other people appearing in it) regions.


\textbf{Obfuscating foreground areas:} Using advanced methods (3D avatars and deepfakes) on the face can be perceived to provide better privacy \emph{(all $\chi^2$(1)>29, p<0.001)} and fairness \emph{(all $\chi^2$(1)>44, p<0.001)} than applying conventional blurring. However, although 3D avatars could offer better privacy \emph{($\chi^2$(1)=313.71, p<0.001)} and fairness \emph{($\chi^2$(1)=100.30, p<0.001)} compared to deepfakes, they might hide more cheating information \emph{($\chi^2$(1)=180.81, p<0.001)} than the latter. Interestingly, a deepfake with a changed skin tone compared to the original skin tone may not significantly impact the dependent variables \emph{(all $\chi^2$(1)>0.02, p>0.57)}. On the other hand, when obfuscating body parts, using deepfake to replace it with a covered outfit may be better on all dependent variables than merely blurring them \emph{(all $\chi^2$(1)>17, p<0.001)}.

\textbf{Obfuscating background areas:} Of the two expert-recommended methods (blurring and deepfake) for obscuring the static background, using deepfake to replace it with an image can be perceived as a better method for both privacy \emph{($\chi^2$(1)=467.86, p<0.001)} and fairness \emph{($\chi^2$(1)=309.18, p<0.001)} - than just blurring it. But the former may hide more cheating information than the latter \emph{($\chi^2$(1)=65.39, p<0.001)}. Also, if people show up in the background, replacing them with a silhouette-like figure can be found to be most effective for both privacy \emph{(all $\chi^2$(1)>95, p<0.001)} and fairness \emph{(all $\chi^2$(1)>79, p<0.001)}, better than using either a full-body 3D avatar or blurring. Yet, using the silhouette could compromise on retaining cheating information more than either a full-body 3D avatar or blurring \emph{(all $\chi^2$(1)>116, p<0.001)}. Based on the analysis conducted so far, no single obfuscation method emerged as the best choice across all three dependent variables in different ROIs. Therefore, we will proceed to examine the effects of obfuscation methods on the overall perception.

\subsubsection{Effects of Region-specific Obfuscation Methods on the Perceptions Combined} \label{sec:combined}

To address RQ5, which examines the impact of obfuscation methods on participants' overall perceptions, we summed the ratings of all three dependent variables. Considering the variables were significantly correlated to each other (most have r>0.15, p<0.001; see Appendix Table~\ref{tab:corr}), composite scores were only used to measure the overall perception without encompassing a broader concept. PANEL 2 in Figure~\ref{fig:allthree} presents these composite scores compared to the new composite baselines. We ran four mixed-effects models followed by post hoc analyses (details in Appendix Table~\ref{tab:statistics} and \ref{tab:waldresult}). In all models, variability was low (<3.24) on the new scale across ROIs, indicating that obfuscation methods had a greater impact than individual differences.


\textbf{Obfuscating foreground areas:} 
When obscuring test-takers' face and body, the overall perceptions of the obfuscation method's effectiveness can be influenced more by perceived privacy \emph{(all r>0.67, p<0.001)} and fairness \emph{(all r>0.71, p<0.001)}, than by the amount of cheating information being suppressed \emph{(all r>0.16, p<0.001)}. Using a 3D avatar for the face can perform significantly better overall than both deepfake and blurring \emph{(all $\chi^2$(1)>136, p<0.001)}. Changing the face color with deepfake may not significantly alter perceptions \emph{($\chi^2$(1)=0.13, p=0.72)}. For body obfuscation, using deepfake to change the outfit can significantly provide better overall perception than just blurring it \emph{($\chi^2$(1)=150.45, p<0.001)}.

\textbf{Obfuscating background areas:} 
Similar to foreground areas, when we obscure background elements, test-takers may assess the methods' effectiveness more based on their perceptions of privacy \emph{(all r>0.64, p<0.001)} and fairness \emph{(all r>0.76, p<0.001)} rather than the extent of cheating information suppression \emph{(all r>0.03, p<0.001)}. Using deepfake to replace a static background can produce better results compared to merely blurring it \emph{($\chi^2$(1)=448.93, p<0.001)}. Additionally, replacing individuals in the background, if present, with a silhouette-like figure can prove more effective than employing either a 3D avatar or blurring techniques \emph{(all $\chi^2$(1)>17, p<0.001)}. Having identified which obfuscation methods work best for specific ROIs based on the overall perceptions, we'll next explore if participants' UX aligns with those perceptions.


\subsubsection{Effects of Region-specific Obfuscation on Willingness to Share Obfuscated Video}

PANEL 3 in Figure~\ref{fig:allthree} addresses RQ6 by showing average willingness ratings for each region-specific obfuscation method. We used four mixed-effects models for four ROIs, followed by post hoc analyses. Since the willingness was not assessed in the no-obfuscation condition, the models considered ratings for blurring as the baseline due to its conventional use. Detailed results are in Appendix Table~\ref{tab:statistics} and \ref{tab:waldresult}. Variability was moderately low across all ROIs \emph{(<2.09)}, indicating that obfuscation methods had a greater impact on the willingness ratings than participants' individual differences. Key findings are reported below.

In the face region, blurring was significantly more preferred than either deepfakes or 3D avatars \emph{(all $\chi^2$(1)>44, p<0.001)}. Interestingly, participants were much less willing to share their videos when deepfaked with skin color change compared to without skin color change \emph{($\chi^2$(1)=448.93, p<0.001)}. A similar preference for blurring was found for body and background individuals, except in the background where both blurring and deepfake had high ratings with no significant difference \emph{($\chi^2$(1)=1.03, p=0.31)}. Contrary to the findings in Section~\ref{sec:combined}, where participants' overall perceptions favored advanced obfuscation methods (e.g., 3D avatar, deepfake), they predominantly opted for blurring in all ROIs for sharing videos with reviewers. Since the overall perceptions didn't match their sharing preferences, we checked for correlations and found either no relationship for foreground areas (face and body) \emph{(all r<0.04, p>0.45)} or low correlation \emph{(all r<0.2, p<0.001)} for background areas (still background and people). This will be discussed further in the discussion section.


\subsubsection{Qualitative Analysis of Scope for Discrimination Post-Obfuscation} \label{sec:qualitative}
Our study design accounted for potential discriminatory factors (skin tone, ethnicity and gender) based on expert recommendations while creating test stimuli. Hence, we analyzed the open-ended questions using an inductive coding process in MAXQDA (v. 2024) to understand participants' perceptions of obfuscation addressing these factors. The first author generated codes during the inductive coding process, given the short and straightforward nature of the data (codes are provided in Appendix Table~\ref{tab:qualitativecode}).

We have already observed in Section~\ref{sec:individual_effects} and \ref{sec:combined} that the alteration of participants' skin tone while using the deepfake method didn't show significant effects. However, more participants (around one-third) expressed concerns in open-ended responses regarding the alteration of a fair-skinned test-taker to a darker skin color, compared to one-fifth in the reverse scenario. They cited that altered skin tone could bias reviewers' perceptions of assumed ethnicity. Next, we didn't statistically compare the obfuscation method's effectiveness across genders due to simultaneous alterations in the test stimuli beyond gender, e.g., body, background. However, open-ended responses highlighted persistent cues implying the assumed gender of test-takers despite obfuscation in certain ROIs. Concerns often centered around unchanged length and textures of hair in the deepfaked face; gender-specific outfits chosen for body replacement using deepfake; and differences in body builds even after replacing the background individuals with a silhouette. Using 3D avatars for replacing test-takers' faces or background individuals' full bodies also raised concerns regarding the selected avatar color and gender alignment, particularly when background individuals were replaced with avatars of the same gender.

Fairness concerns were also reported beyond biases related to skin tone, ethnicity and gender. For instance, inferring test-takers' socio-economic status from unconventional test-taking places (e.g., a warehouse used in our study design) or perceiving the presence of background individuals during test-taking as unprofessional could potentially lead to discrimination. Other concerns arose like potential distraction caused by visually appealing 3D avatars for background individuals. Furthermore, suspicions of cheating may arise from complete background replacement with a picture or substituting background individuals with a silhouette, potentially undermining the fairness of obfuscation measures.

\section{Discussion}
This paper addresses key challenges in remote proctoring that may affect test-taking experience. By exploring promising obfuscation methods that can hide privacy-sensitive details in video recordings, the study aims to improve test-takers' experience, for example, their perceptions on privacy protection. It further explores whether these obfuscation methods can eliminate potential discriminatory attributes in the videos (e.g., test-takers' ethnicity, gender) \cite{yoder2022racial}, which may unfairly influence reviewers' judgments and lead to unjust accusations of cheating. Finally, these methods must not compromise the core purpose of remote proctoring, which is to provide reviewers with videos containing enough information to detect cheating. Test-takers' perception of this information sufficiency can influence their test-taking experience and their likelihood of cheating, assuming they believe that cheating actions can be identified.

In this study, we interviewed experts from e-assessment, computer vision and usable privacy fields to identify the most promising obfuscation methods applicable to various regions of a video, such as test-takers' face, body, background and people in the background. Subsequently, we examined how these methods might affect non-experts' (acting as potential test-takers) perceptions of privacy protection, fairness and information sufficiency, as well as their willingness to adopt each of those methods for remote proctoring.

We explored obfuscation methods with distinct working principles, potentially evoking varied user perceptions \cite{li2017effectiveness}. For instance, blurring gradually fades content like faces; deepfake merges content characteristics with a reference to create a new realistic output; 3D avatar overlays a reference to fully conceal the content; and silhouette replacement blacks out content based on its contour, potentially revealing the original content due to the visible outline.

Acknowledging the computational limitations of implementing promising methods like deepfake in videos, we simplified our user testing by using simulated images instead. Participants were instructed to imagine these images as visual scenes in video recordings. While static images may not fully capture the dynamic nature of video content, they can still encapsulate crucial elements influencing user perceptions \cite{roberts2009static, conway2008evidence}. As these promising obfuscation techniques mature, they may become more viable for integration into video editing workflows in the future. Below, we provide insights on obfuscations, contributing to the discourse surrounding standardizing video recording in remote proctoring.

\subsection{Obfuscating Foreground Areas of Remote Proctored Videos} \label{sec:foreground}
Previous studies \cite{li2017effectiveness, chen2009protecting} found that obscuring both the face and body of test-takers provides greater privacy protection than just obscuring the face. However, in our study, we did not uniformly obfuscate the full bodies of test-takers. Instead, we evaluated different obfuscation methods separately for the face and body because different regions of a video may offer different privacy-utility trade-offs \cite{ravi2023review, erdelyi2018privacy}. For example, some regions may have high privacy significance, while others may be most informative for cheating detection (utility). Additionally, our study considers fairness in cheating detection as another important factor in this trade-off, alongside privacy and cheating detection.

The test-taker's face is highly identifiable, raising significant privacy and fairness concerns. Obfuscating the face is challenging because facial expressions, head pose and eye gaze are crucial for detecting cheating behavior. While deepfake with realistic face replacement can commonly be used for this purpose \cite{tolosana2020deepfakes}, our evaluation by potential test-takers (in Section~\ref{sec:combined}) indicated that a cartoonized 3D avatar replacement could be more effective, based on the overall perceptions of privacy, fairness and information sufficiency. This relative effectiveness of 3D avatars reflected in higher privacy and fairness ratings (see Figure~\ref{fig:allthree}), aligns with a prior study \cite{li2017effectiveness} indicating similar privacy benefits with avatarized face representations. However, it's unclear if this effect was due to the obfuscation methods themselves or subtle differences between their designs. For example, during stimuli design, the gender-neutral cartoon face used in 3D avatars contrasts with deepfake methods that retain the test-takers' original facial characteristics (e.g., hair), potentially prompting assumptions about their gender or ethnicity and raising fairness concerns. Additionally, the adjustable size of 3D avatars can hide further cues, such as neck color, thus mitigating assumptions about the test-taker's ethnicity.

Body obfuscation is also important because the body can reveal sensitive cues such as attire type or tattoos \cite{coghlan2021good, kharbat2021proctored}. The best way to conceal these cues might involve replacing the entire body region with a realistically covered outfit. Deepfake is a promising candidate for this, as it preserves motion information \cite{ravi2023review}, such as frequent hand or shoulder movements, crucial for cheating detection. Collectively, our findings imply that achieving an optimal level of obfuscation in remote proctoring videos may entail applying distinct obfuscation techniques for the face and body regions, while simultaneously ensuring the preservation of critical motion information essential for detecting cheating behaviors.

\subsection{Obfuscating Background Areas of Remote Proctored Videos} 
Obfuscation of background areas poses a unique challenge as it involves concealing sensitive visual cues such as living conditions, specific objects or the presence of other individuals (e.g., family members) \cite{terpstra2023online} without losing relevant information that may assist test-takers. Background obfuscation requires a distinct strategy for static backgrounds and individuals appearing in the background. Unlike the separate obfuscation for the test-takers' face and body, as discussed in Section~\ref{sec:foreground}, the lower frequency of individuals appearing in the background may allow for a full-body obfuscation.

A promising approach could involve replacing the static background with a generic picture for all test-takers using deepfake, similar to video conferencing tools, while excluding any individuals detected in the background. This also mitigates potential discrimination against test-takers based on factors like social status or lack of professional background, especially when tests are taken in unconventional places, as discussed in Section~\ref{sec:qualitative}. Next, adhering strictly to remote testing guidelines for an ideal test-taking environment without visible individuals may be impractical for test-takers with diverse living situations. However, if interactions between test-takers and those individuals are strictly prohibited during test-taking, a silhouette-like figure replacing their entire body, while preserving motion and contour, can still be informative for cheating detection while ensuring privacy protection. However, caution is needed, as the silhouette's contour may lead to assumptions about their gender, potentially undermining fairness expectations.

\subsection{Visual Obfuscation - a Preferred Solution for Remote Proctoring?} 
State-of-the-art obfuscation methods (e.g., deepfake, 3D avatar) appear promising in offering test-takers adequate levels of perceived privacy protection, perceived fairness and perceived information sufficiency. However, non-expert participants expressed hesitancy in sharing their videos with unknown reviewers, particularly when obfuscating their faces or bodies using these methods. In contrast, blurring emerged as the preferred solution for both regions compared to those advanced methods. This resonates with the findings from a prior study \cite{li2017effectiveness} on applying obfuscation methods to social media photos, where solutions offering lower privacy but higher information sufficiency were favored. 
Our study also considered fairness aspects alongside privacy and cheating detection in the context-dependent evaluation of the privacy-utility trade-off \cite{valdez2019users}. Based on the willingness ratings, it appears that the perceived effectiveness of those advanced techniques (i.e., deepfake, 3D avatar) in retaining crucial cheating information may be lacking, or there might be a lack of trust \cite{dwyer2007trust, schudy2017you} in their practical implementation. Test-takers may also have concerns about potential video glitches, issues related to avatar design, or the selection of gender and skin tone for face or body replacement, that can affect obfuscation effectiveness. Blurring may also be preferred due to its familiarity and cost-effectiveness. Hence, addressing these concerns could bolster test-takers' trust in more advanced obfuscation methods and encourage them to share obfuscated videos more willingly. Taken together, our study offers insights into the expected outcomes when implementing obfuscation techniques on video recordings, paving the way for further research, as outlined in Section~\ref{sec:limitation}.

\label{sec:discussion}

\section{Actionable Guidelines for Researchers and Practitioners}

Our study suggests insights with following guidelines for balancing test-takers' perceptions of privacy and fair cheating detection in remote proctoring setups, particularly if obfuscation solutions are applied during post-test video processing.

    \textbf{(1) Effective obfuscation methods for distinct regions}:
    
    \textbf{\emph{A. Foreground region}}: Replacing a test-taker's face with a sufficiently large uniform 3D avatar face can effectively hide facial features, including the neck, while retaining real-time expressions like eye gaze, mouth and head movements, crucial for successful cheating detection. Next, a standardized gender-neutral professional outfit, such as a formal suit, can effectively replace the body region, preserving the movement of hands and shoulders crucial for cheating detection.
    
    \textbf{\emph{B. Background region}}: Choosing a uniform generic background image for replacement can effectively conceal test-taking environment and bias-inducing elements, but test-organizing institutes should verify for presence of cameras behind test-takers in unprocessed videos after reviewers assess the obfuscated videos, crucial for cheating detection. Individuals in the background can be replaced with a mono-colored 2D silhouette-like figure by accurately measuring their body contour and retaining real-time body movements.

    \textbf{(2) Practical remote proctoring pipeline}: A hybrid remote proctoring approach combines algorithmic-based cheating detection with manual verification. Recorded videos should be securely stored with a trusted entity; relevant obfuscation methods can be applied to a copy of the recorded videos and shared with external reviewers hired by test-organizing institutes (refer to Figure~\ref{fig:pipeline}). Only the institutes would have access to the unprocessed videos for a final review based on reports from reviewers, to resolve potential disputes if test-takers challenge cheating allegations.
\label{sec:recommendation}

\section{Limitation and Future Scope}
While this study highlights the potential of region-specific obfuscation methods to improve the test-taking experiences, future research could expand upon this work to fully understand and realize their benefits in remote proctoring.

\emph{First}, while our survey relied on static images to illustrate obfuscation methods, future studies could apply these methods directly to video recordings of test-takers to provide participants with a more realistic portrayal. However, this approach would require addressing various technical, logistical, legal and ethical challenges. 
\emph{Second}, exploring the perspectives of proctoring managers could offer valuable insights into the practical challenges and feasibility of implementing obfuscation techniques in real-world scenarios. 
\emph{Third}, we constructed scales (Table~\ref{tab:measurements}) for test-takers' perception of privacy, cheating information sufficiency and fairness using ad-hoc items. The development and validation of standardized scales for these dimensions could greatly benefit researchers in this area. 
\emph{Finally}, while our survey sampled a diverse group of participants as potential test-takers, future studies could target specific populations such as students or professionals to explore potential differences in their perceptions and experiences with remote proctoring.

\label{sec:limitation}

\section{Conclusion}

This study marks an initial step toward enhancing test-takers' experiences in remote proctoring by addressing key concerns surrounding the review of proctored videos: protecting their privacy and accurate and fair cheating detection. Through the selective obfuscation of privacy-sensitive visual data in various video regions (test-taker's face, body, background and individuals in the background) using expert-recommended (N=9) obfuscation methods, we evaluated their impact on potential test-takers' (N=259) overall perceptions of those three dimensions as well as their willingness to share obfuscated videos. Our findings underscore the significant impact of region-specific obfuscation methods on participant experiences, suggesting that optimal outcomes may be achieved by tailoring obfuscation methods across regions (e.g., 3D avatar on the face, deepfake on the body, silhouette on background individuals). We provide guidance for researchers and practitioners to assess the cost-effectiveness of testing these methods with real proctored videos before practical implementation, while also advocating for further research in this pertinent and evolving field.

\label{sec:conclusion}

\bibliographystyle{ACM-Reference-Format}
\bibliography{10_bibliography}
\break
\appendix
\section{Materials used in Expert Evaluation}

\subsection{Recruitment of Experts for Interviews}
\begin{table}[H]
  \scriptsize
  \caption{Details of experts being interviewed}
  \label{tab:experts-profile}
  \begin{tabular}{lll}
    \hline
    \textbf{\makecell[c]{Experts}} & \textbf{\makecell[c]{Designation}} & \textbf{\makecell[c]{Expertise}} \\
    \hline
    
    \multirow{4}{*}{\makecell[l]{E-Assessment\\ Experts}} 
    & \multirow{1}{*}{\makecell[l]{Programme leader}} & 4+ years in proctoring university exams and privacy related research \\
    \cmidrule(lr){2-3}
    & \multirow{1}{*}{\makecell[l]{Programme manager}} & $\sim$5 years in proctoring university exams \\
    \cmidrule(lr){2-3}
    & \multirow{1}{*}{\makecell[l]{Assessment specialist}} & 1+ year in proctoring university exams  \\
    \hline 
    \multirow{5}{*}{\makecell[l]{Computer Vision \\ Experts}} 
    & \multirow{1}{*}{\makecell[l]{Key researcher}} & 20+ years in CV research, with 10+ years particularly in surveillance domain \\
    \cmidrule(lr){2-3}
    & \multirow{1}{*}{\makecell[l]{Delivery head}} & 10+ years in the surveillance domain with 5 years in privacy protection research\\
    \cmidrule(lr){2-3}
    & \multirow{1}{*}{\makecell[l]{Staff research scientist}} & 4 years in digital face manipulation research using neural networks \\
    \cmidrule(lr){2-3}
    & \multirow{1}{*}{\makecell[l]{Scientific staff}} & 3 years in CV research with biometric data protection \\
    \hline
    \multirow{3}{*}{\makecell[l]{Privacy \\ Experts}} 
    & \multirow{1}{*}{\makecell[l]{Assistant professor}} & 5 years in usable privacy research in remote proctoring \\
    \cmidrule(lr){2-3}
    & \multirow{1}{*}{\makecell[l]{Associate professor}} & 8 years in HCI research, specializing in usable privacy for ubiquitous systems\\
    \hline
  \end{tabular}
\end{table}

\subsection{Tasks During Experts' Interviews}
\begin{figure}[H]
  \centering
  \includegraphics[width=0.6\textwidth]{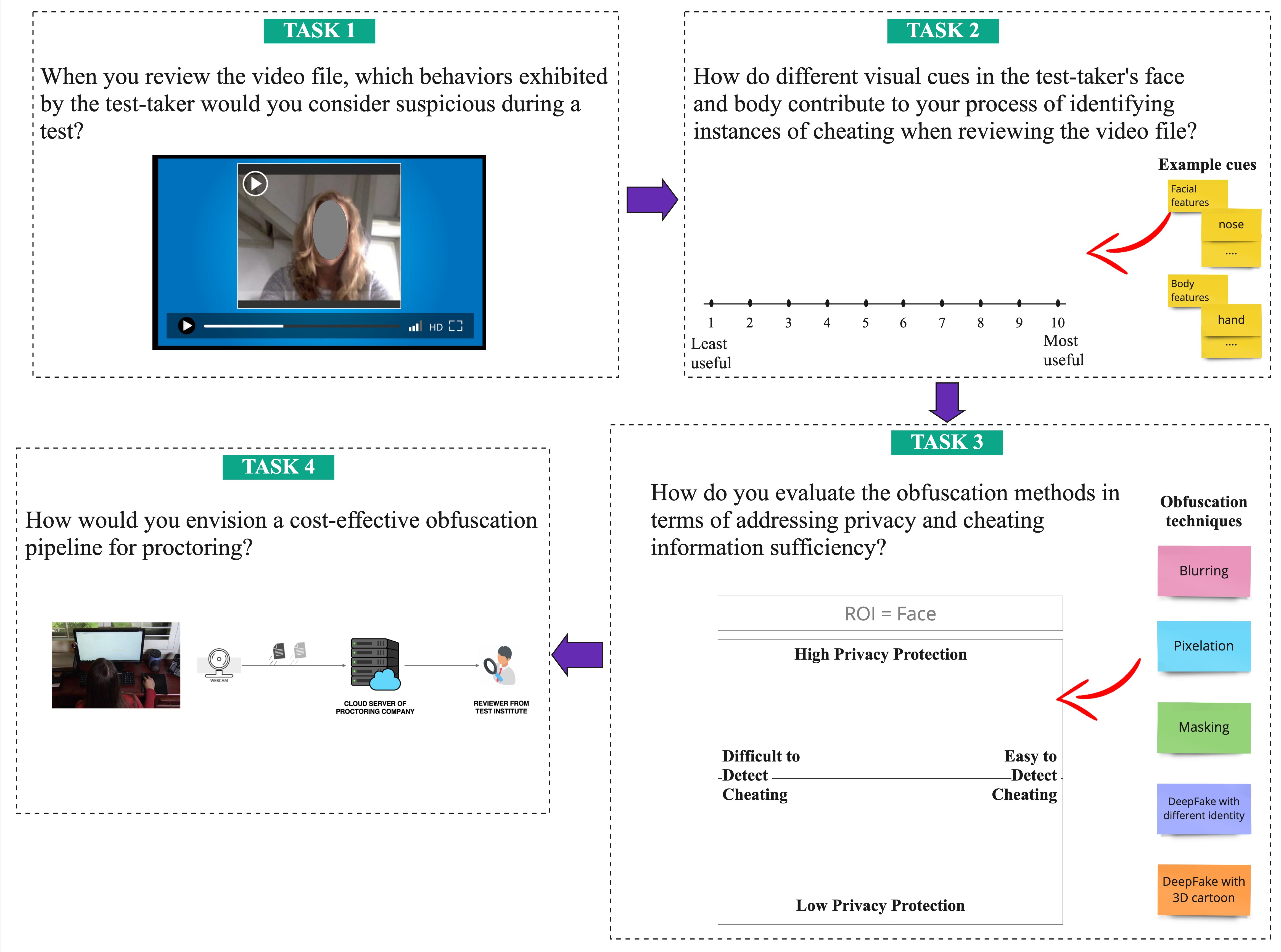}
  \caption{The expert interviews followed a sequence of tasks: \textit{Task 1} involved listing cheating instances for each region in the image shown; \textit{Task 2} focused on identifying and rating (1-10) visual information that could be suppressed through obfuscation without compromising cheating detection. Example only shows cues for face and body. Other regions were also discussed; \textit{Task 3} allowed experts to assess the privacy-cheating trade-off for each method in each region. They were required to drag and drop according to their assessment. \textit{Task 4} involved a discussion on the obfuscation pipeline, guided by the existing flow diagram of remote proctoring}
\label{fig:task}
\end{figure}

\subsection{Codebook for Qualitative Analysis of Expert Interviews} \label{appdx:qualcodes}
\begin{enumerate}

    \item \textbf{Identify cheating behaviors}: This category has 4 subcategories based on 4 video regions or ROIs.
    \begin{itemize}
        \item 1.1 Cheating behaviors in face
        \item 1.2 Cheating behaviors in body
        \item 1.3 Cheating behaviors in background
        \item 1.4 Cheating behaviors in people in the background
    \end{itemize}
    The codes are repetitive for all subcategories, mentioned with example quotes. They are: 1) \textbf{dishonest behaviors} \textit{("...it's clear that that this person is talking to someone else")}, 2) \textbf{unauthorized behaviors} \textit{("obviously if another person appears on the screen..., I assume the prerequisite for this exam was to be alone in the room"}), 3) \textbf{unusual behavior pattern} \textit{("If the shoulder is moving..., the hands are moving, more than needed for typing ends")} and 4) \textbf{risk of false flags} \textit{("frozen screen and stuff like that. And so the things can go wrong there...")}
    
    \item \textbf{Useful visual cues for cheating detection}: This category also has 4 subcategories based on 4 video regions or ROIs.  
    \begin{itemize}
        \item 1.1 Visual cues in face
        \item 1.2 Visual cues in body
        \item 1.3 Visual cues in background
        \item 1.4 Visual cues in people in the background
    \end{itemize}
    Repetitive codes for all subcategories are mentioned with example quotes: 1) \textbf{crucial cues to preserve} \textit{("face pose is a logical indicator, or say meaningful to understand where she's looking")}, 2) \textbf{traits can be concealed} \textit{("not sure umm...how she looks adds any value to cheating, all it matters to form facial expressions ...that can be helpful to have"}), and 3) \textbf{cues as frequent indicators} \textit{("depends on cheating, I think, hmm. most visible part is face ... you can have idea looking at her face if she's cooking something")}
    
    \item \textbf{Evaluation of obfuscation methods}: This category also has 4 subcategories based on 4 video regions or ROIs.
    \begin{itemize}
        \item 1.1 Evaluation with face
        \item 1.2 Evaluation with body
        \item 1.3 Evaluation with background
        \item 1.4 Evaluation with people in the background
    \end{itemize}
    Repetitive codes for all subcategories are mentioned with example quotes: 1) \textbf{privacy evaluation} \textit{("...blurring is actually low privacy and the actual subject face can be retained with a different filter.. they can see the body and the face, and the colour, probably gender...")}, and 2) \textbf{cheating detection evaluation} \textit{("I see you want to change the outfits. Replacement is a good technique in my opinion for privacy, how about they hide cheat sheet in their.. Let's say shirt's pockets. You miss out on important information for your cheating assessment"})
    
    \item \textbf{Ensure fairness in cheating detection}: This category has 2 subcategories.
    \begin{itemize}
        \item 1.1 Measures for test-takers: The codes: 1) \textbf{gender} \textit{("Doesn't matter if you're woman or a man. Same face for everyone if you are replacing their faces... But Very boring as a reviewer to watch them having the similar face. ok. Not sure about any new bias for it. ...People always try to make an assumption.")}, and 2) \textbf{skincolor} \textit{("I would also fix the colour for also to avoid biases when I will put a. Let's say 3D fictional face. I would just keep it like a standard colour umm or something you know so it doesn't give out"})
        \item 1.2 Measures for people in the background: Only one code: \textbf{gender} \textit{("...I would prefer virtual avatars. I see. for the full body. It should be same avatar for all ...Keeping gender constant")}
    \end{itemize}
    
    \item \textbf{Practical obfuscation pipeline} This category has two codes: 1) \textbf{access to original video} \textit{("The first line of review, the Proctor slash reviewers. They will, they will watch some sort of Anonymized version. Can still detect all the important features. if the secretaries of the examination boards in our situation would watch the normal video without anonymization.")}, and 2) \textbf{issues with obfuscated videos} \textit{("I think students would feel much more comfortable. Secretaries can still check if you are wrongly flagged I think. ...huge issues when you rely only on Anonymized versions. Solving privacy issues umm introducing ethical issues even more, some trust issues")} 
\end{enumerate}

\section{Materials used in Non-expert Evaluation}

\subsection{Details of Items used in Survey Questionnaire}
\begin{table}[h]
  \scriptsize
  \caption{Measurements during vignette experiment with non-experts}
  \label{tab:measurements}
  \begin{tabular}{lll}
    \hline
    \textbf{Scale} & {\makecell[l]{\textbf{Region of}\\ \textbf{Interest (ROI)}}} & \textbf{\makecell[c]{Items used in the survey with 7-point Likert scale on agreement}} \\
    \hline
    
    \multirow{9}{*}{\makecell[l]{Perceived \\ information\\ sufficiency}} &  \multirow{2}{*}{\makecell[l]{Face}} & 1. Reviewers could detect when this test-taker carries out mouth movement, such as talking. \\
    && 2. Reviewers could detect when this test-taker interacts with other people (in the room).\\
    \cmidrule(lr){2-3}   
    & \multirow{2}{*}{\makecell[l]{Body}} &  1. Reviewers could detect test-taker's body movements suggesting unauthorized use of resources (e.g smartphone). \\
    && 2. Reviewers could detect when the test-taker moves from their seat.\\ 
    \cmidrule(lr){2-3}
    & \multirow{2}{*}{\makecell[l]{Background}} & 1. Reviewers could detect when the test-taker interacts with other people in their environment.   \\
    && 2. Reviewers could detect the presence of a camera in the background recording the test-taker's computer screen.\\
    \cmidrule(lr){2-3}
    & \multirow{2}{*}{\makecell[l]{People in background}} & 1. Reviewers could detect when the person in the background talks.  \\
    && 2. Reviewers could detect when the test-taker interacts with the person in the background. \\
    \hline

    \multirow{6}{*}{\makecell[l]{Perceived\\ privacy\\ concerns}} 
    & \multirow{1}{*}{\makecell[l]{Face}} & 1. Reviewers could recognize the identity of the test-taker. \\
    \cmidrule(lr){2-3}
    
    & \multirow{2}{*}{\makecell[l]{Body}} & 1. Reviewers could recognize the outfits of the test-taker. \\
    && 2. Reviewers could recognize the body identity of the test-taker.\\ 
    \cmidrule(lr){2-3}
    & \multirow{1}{*}{\makecell[l]{Background}} & 1. Reviewers could recognize the objects present in the background.  \\
    \cmidrule(lr){2-3}
    & \multirow{1}{*}{\makecell[l]{People in background}} & 1. Reviewers could recognize the identity of the person in the background.  \\
    \hline
     
    \multirow{3}{*}{\makecell[l]{Perceived \\ fairness\\ concerns}} & \multirow{2}{*}{\makecell[l]{Face, Body,\\ Background,\\ People in background}} & 1. Biased reviewers could exhibit discrimination against this test-taker.  \\
    \cmidrule(lr){3-3}
    && \makecell[l]{(Open-ended question) What aspect of this test-taker's modified image might be used by a biased reviewer to \\discriminate against this test-taker?} \\
    \hline

    \multirow{3}{*}{\makecell[l]{Willingness to \\ video sharing}} & \multirow{3}{*}{\makecell[l]{Face, Body,\\ Background,\\ People in background}} &  \\ &&1. I am willing to share my modified video with the reviewers.\\&&\\
    \hline
  \end{tabular}
\end{table}

\subsection{Correlation between Dependent Variables used in Non-expert Evaluation} \label{appdx:corr}

\begin{table}[H]
\caption{Pairwise Correlations between dependent variables. Pearson's correlation coefficient. $^+$p<0.1, $^*$p<0.05, $^{**}$p<0.01, $^{***}$p<0.001}
\label{tab:corr}
\scriptsize
\setlength\tabcolsep{0pt}
\begin{tabular*}{\linewidth}{@{\extracolsep{\fill}}
                                l*{5}{c}
                            }
    \toprule
      \textbf{ROI : Face}  & (1)   & (2)   & (3) & (4) & (5)\\
    \midrule
(1): Perceived privacy    &1.00 &&&\\
(2): Perceived info. sufficiency     &-0.29*** &1.00 &&\\
(3): Perceived fairness &0.31*** &-0.15*** &1.00&\\
(4): Combined ratings   &0.66*** &0.31*** &0.69*** &1.00\\ 
(5): Willingness to video sharing   &-0.02 &-0.01 &0.01 &-0.01 &1.00\\
    \toprule
      \textbf{ROI : Body}  & (1)   & (2)   & (3) & (4) & (5)\\
    \midrule
(1): Perceived privacy    &1.00 &&&\\
(2): Perceived info. sufficiency     &-0.25*** &1.00 &&\\
(3): Perceived fairness &0.24*** &-0.06$^+$ &1.00&\\
(4): Combined ratings   &0.61*** &0.34*** &0.79*** &1.00\\ 
(5): Willingness to video sharing   &-0.13** &0.18 &0.05 &0.03 &1.00\\
    \toprule
      \textbf{ROI : Background}  & (1)   & (2)   & (3) & (4) & (5)\\
    \midrule
(1): Perceived privacy    &1.00 &&&\\
(2): Perceived info. sufficiency     &-0.44*** &1.00 &&\\
(3): Perceived fairness &0.64*** &-0.34*** &1.00&\\
(4): Combined ratings   &0.78*** &0.01 &0.84*** &1.00\\ 
(5): Willingness to video sharing   &0.07$^+$ &-0.01 &0.19*** &0.16*** &1.00\\
    \toprule
      \textbf{ROI : People in background}  & (1)   & (2)   & (3) & (4) & (5)\\
    \midrule
(1): Perceived privacy    &1.00 &&&\\
(2): Perceived info. sufficiency     &-0.45*** &1.00 &&\\
(3): Perceived fairness &0.37*** &-0.26*** &1.00&\\
(4): Combined ratings   &0.64*** &0.09** &0.77*** &1.00\\ 
(5): Willingness to video sharing   &0.16*** &-0.07$^+$ &0.18*** &0.19*** &1.00\\
    \bottomrule
\end{tabular*}
\end{table}

\subsection{Code Systems for Non-expert Evaluation}
\begin{table}[H]
  \scriptsize
   \caption{Code systems of qualitative analysis of fairness perception}
  \label{tab:qualitativecode}
  \begin{tabular*}{\textwidth}{@{\extracolsep{\fill}}llllll@{}}
 
    \hline
    \textbf{Categories} & {\makecell[l]{\textbf{Codes}}} & {\makecell[l]{\textbf{Subcodes}}} & \textbf{\makecell[c]{Description of codes}} & {\makecell[l]{\textbf{Concerned obfuscated regions}}}\\
    \hline
    
    \multirow{16}{*}{\makecell[l]{Biases based \\on test-takers'\\ characteristics}} 
    & \multirow{1}{*}{\makecell[l]{Gender}} & &{\makecell[l]{Participants believed that the gender of the subject could still \\ be inferred from obfuscated regions}}& Face, body, people in background\\
    \cmidrule(lr){2-5}
    & & {\makecell[l]{Gender\textbackslash Body shape}} &{\makecell[l]{Gender was believed to be inferred from displayed body shape}}& Body, people in background\\
    \cmidrule(lr){2-5}
    & &{\makecell[l]{Gender\textbackslash Attire}} &{\makecell[l]{Gender was believed to be inferred from displayed attire}}& Body, people in background\\
    \cmidrule(lr){2-5}
    & &{\makecell[l]{Gender\textbackslash Hair}} &{\makecell[l]{Gender was believed to be inferred from displayed hair length}}& Face, people in background\\
    \cmidrule(lr){2-5}
    & \multirow{1}{*}{\makecell[l]{Ethnicity}} & &{\makecell[l]{Participants believed that the ethnicity of the subject could still\\ be inferred from obfuscated regions}} &Face, body, people in background\\
    \cmidrule(lr){2-5}
    & & {\makecell[l]{Ethnicity\textbackslash Skin tone}} &{\makecell[l]{Ethnicity was believed to be inferred from displayed skin tone}} &Face, body, people in background\\
    \cmidrule(lr){2-5}
    & &{\makecell[l]{Ethnicity\textbackslash Attire}} &{\makecell[l]{Ethnicity was believed to be inferred from displayed attire}}& Body, people in background\\
    \cmidrule(lr){2-5}
    & &{\makecell[l]{Ethnicity\textbackslash Hair}} &{\makecell[l]{Ethnicity was believed to be inferred from displayed hair color, type}}& Body, people in background\\
    \cmidrule(lr){2-5}
    & \multirow{1}{*}{\makecell[l]{Social status}} & &{\makecell[l]{Participants believed that the social status of the subject could still\\ be inferred from obfuscated regions}} &{\makecell[l]{Body, background,\\ people in background}}\\
    \cmidrule(lr){2-5}
    &  & {\makecell[l]{Social status\textbackslash Background}}&{\makecell[l]{Social status was believed to be inferred from displayed background}} &{\makecell[l]{Background, people in background}}\\
    \cmidrule(lr){2-5}
    &  & {\makecell[l]{Social status\textbackslash Attire}}&{\makecell[l]{Social status was believed to be inferred from displayed attire}} &{\makecell[l]{Body, people in background}}\\
    
    \hline
    
    \multirow{6}{*}{\makecell[l]{Biases based \\on other\\ aspects}} 
    & \multirow{1}{*}{\makecell[l]{Unprofessionalism}} & Test-taking environment&{\makecell[l]{Participants mentioned unusual test-taking place and people \\appearing in the background as unprofessional}}&Background, people in background\\   
    \cmidrule(lr){2-5}
    & \multirow{1}{*}{\makecell[l]{Distraction}} & &{\makecell[l]{Participants believed unusual test-taking place and presence of \\others as distracting during review}} &{\makecell[l]{Background, people in background}}\\
    \cmidrule(lr){2-5}
    &  & {\makecell[l]{Distraction\textbackslash Obfuscation \\design}}& {\makecell[l]{Attractive design and presence of technical glitch while obfuscation \\is applied could cause distraction from reviewing}} &{\makecell[l]{Face, body, background,\\ people in background}}\\
    \hline
  \end{tabular*}
\end{table}

\subsection{Pilot Studies for Non-expert Evaluation} \label{appdx:pilot} While modifying the created stimuli by blurring different regions of interest (ROI) e.g., face, body, background and people in the background, a brief pilot test involving six participants was conducted. They adjusted the radius of the blurred ROI using a slider (ranging from 1-100), aiming for a balance between effective ROI concealment and visual information sufficiency for cheating detection. The median value of the collected blurred radius for each ROI was applied to the modified stimuli. We also conducted another pilot study of the entire survey with 10 participants recruited from Prolific, using a feedback box to identify any issues with the questions or image quality. Since no changes were made to the survey afterward, the responses from this pilot test were included in the data analysis.

\subsection{Regression Tables from Non-expert Evaluation}\label{appdx:dep-regression}

\begin{table}[h]
\begin{minipage}{0.45\textwidth}
\scriptsize
    \centering
    \rotatebox{90}{       
    \begin{minipage}{\textheight}
    \caption{Effects of different region-specific obfuscation methods on perceived information sufficiency, privacy protection and fairness. For each mixed-effect model, participant variance, unexplained variance and model fit are also reported}
    \label{tab:statistics}
    \begin{tabular*}{1\textheight}{@{\extracolsep{\fill}}clcccccc@{}}
        \toprule
        & & \multicolumn{2}{c}{\textbf{Perceived information sufficiency}} & \multicolumn{2}{c}{\textbf{Perceived privacy protection}} & \multicolumn{2}{c}{\textbf{Perceived fairness}} \\
        \cmidrule(lr){3-4} \cmidrule(lr){5-6} \cmidrule(lr){7-8} 
        {\makecell{\textbf{Region of} \\ \textbf{interest}}} & {\makecell[l]{\textbf{Obfuscation}\\\textbf{methods}}} & {\makecell{Coeffs with \\ 95\% CI}} & {\makecell{Parameters of\\ mixed-effects model}} & {\makecell{Coeffs with \\ 95\% CI}} & {\makecell{Parameters of\\ mixed-effects model}} & {\makecell{Coeffs with \\ 95\% CI}} & {\makecell{Parameters of\\ mixed-effects model}} \\
        \midrule
        
        \multirow{6}{*}{\makecell{Face of \\ test-taker}} & Blurring  & -1.57*** [-1.71, -1.42] & \multirow{6}{*}{\makecell[l]{Participant var = 0.65 \\ Unexplained var = 0.71 \\ Model fit: $\chi^2$(1)=396.55*** }} & 3.21*** [3.04, 3.37] & \multirow{6}{*}{\makecell[l]{Participant var = 0.48 \\ Unexplained var = 0.89 \\ Model fit: $\chi^2$(1)=220.40*** }} & 0.31*** [0.15, 0.47] & \multirow{6}{*}{\makecell[l]{Participant var = 0.68 \\ Unexplained var = 0.85 \\ Model fit: $\chi^2$(1)=342.89*** }} \\
        
        & Deepfake with$\sim$ & & &&&& \\
        & \hspace{0.2cm}original skin tone   & -0.47*** [-0.62, -0.33] & &  3.67*** [3.51, 3.83] &   & 0.86*** [0.71, 1.02] &  \\
        & \hspace{0.2cm}changed skin tone  & -0.46*** [-0.61, -0.31] & &  3.66*** [3.49, 3.82] & &  0.91*** [0.75, 1.07] &  \\
        
        & 3D avatar  & -1.49*** [-1.63, -1.34] & &  5.16*** [5.01, 5.33]  & & 1.69*** [1.53, 1.85]  &  \\
        
        & No obfuscation & 6.14 & & 1.00&  & 2.25 &  \\
        
        \midrule
        
        \multirow{3}{*}{\makecell{Body of \\ test-taker}} & Blurring  & -0.56*** [-0.66, -0.45] & \multirow{3}{*}{\makecell[l]{Participant var = 0.82 \\ Unexplained var = 0.36 \\ Model fit: $\chi^2$(1)=386.62*** }} & 3.27*** [3.09, 3.44] & \multirow{3}{*}{\makecell[l]{Participant var = 0.17 \\ Unexplained var = 1.07 \\ Model fit: $\chi^2$(1)=13.71*** }}  & 1.15*** [0.64, 1.34] & \multirow{3}{*}{\makecell[l]{Participant var = 0.73 \\ Unexplained var = 1.21 \\ Model fit: $\chi^2$(1)=99.31*** }} \\
        
        & Deepfake  & -0.33*** [-0.43, -0.23] &  & 4.10*** [3.92, 4.28] &  & 1.85*** [1.66, 2.03] &  \\
        
        & No obfuscation & 5.55 & & 1.00&  & 2.46 &  \\
        
        \midrule
        
        \multirow{3}{*}{\makecell{Background \\ of test-taker}} & Blurring  & -1.03*** [-1.19, -0.87] & \multirow{3}{*}{\makecell[l]{Participant var = 0.94 \\ Unexplained var = 0.86 \\ Model fit: $\chi^2$(1)=197.26*** }} & 2.59*** [2.41, 2.77]  & \multirow{3}{*}{\makecell[l]{Participant var = $\sim$0.00 \\ Unexplained var = 1.13 \\ Model fit: $\chi^2$(1)=0.00 }} & -0.09 [-0.31, 0.12] & \multirow{3}{*}{\makecell[l]{Participant var = 0.02 \\ Unexplained var = 1.56 \\ Model fit: $\chi^2$(1)=0.10 }}  \\
        
        & Deepfake  & -1.68*** [-1.84, -1.52] & & 5.06***[4.88, 5.24] &  & 1.85*** [1.63, 2.06] &  \\
        
        & No obfuscation & 5.23 & & 1.00&  & 3.61&  \\
        
        \midrule
        
        \multirow{4}{*}{\makecell{People in \\ background}} & Blurring  & -1.37*** [-1.52, -1.23] & \multirow{4}{*}{\makecell[l]{Participant var = 0.92 \\ Unexplained var = 0.71 \\ Model fit: $\chi^2$(1)=389.21*** }} & 3.62*** [3.43, 3.81] &  \multirow{4}{*}{\makecell[l]{Participant var = 0.68 \\ Unexplained var = 1.15 \\ Model fit: $\chi^2$(1)=166.36*** }} & 0.72*** [0.52, 0.91] &  \multirow{4}{*}{\makecell[l]{Participant var = 1.41 \\ Unexplained var = 1.23 \\ Model fit: $\chi^2$(1)=345.24*** }}\\
        
        & Silhouette  & -2.11*** [-2.25, -1.96] & & 4.97*** [4.78, 5.15] &  & 1.80*** [1.61, 1.99] &  \\
        
        & 3D avatar  & -1.01*** [-1.15, -0.86] & & 4.08*** [3.89, 4.26] &  & 0.97*** [0.78, 1.16] &  \\
        
        & No obfuscation & 5.96 & &  1.00&  & 2.69 &  \\
        
       \hline   
    \multicolumn{8}{l}{- Coefficients of the dependent variables for ROI-specific obfuscation methods are shown relative to the baseline of `no obfuscation' in the mixed-effects models}\\
    \multicolumn{8}{l}{- Mean value index of perceived information sufficiency, perceived privacy and perceived fairness on a scale of 1 to 7}\\
    \multicolumn{8}{l}{- $^+$p<0.1, $^*$p<0.05, $^{**}$p<0.01, $^{***}$p<0.001}\\
    \bottomrule
    \end{tabular*}
    \end{minipage}
    } 
    \end{minipage}
    \hfill
\begin{minipage}{0.45\textwidth}
\scriptsize
    \centering
    \rotatebox{90}{
        \begin{minipage}{\textheight}
    \caption{Comparison of effects of different region-specific obfuscation methods on dependent variables. Deepfake (O) and (C) represent deepfaking with original skin tone and changed skin tone respectively}
    \label{tab:waldresult}
    \begin{tabular*}{1\textheight}{@{\extracolsep{\fill}}cllllll@{}}
        \toprule
        {\makecell{\textbf{Region of} \\ \textbf{interest}}} & {\makecell[l]{\textbf{Comparing effects}\\\textbf{between methods}}} & {\makecell{\textbf{Perceived informa-}\\\textbf{tion sufficiency}}}  & {\makecell{\textbf{Perceived privacy}\\ \textbf{protection}}}  &  {\makecell{\textbf{Perceived}\\ \textbf{fairness}}}  & {\makecell{\textbf{Composite} \\\textbf{scores}}} & {\makecell{\textbf{Willingness to} \\\textbf{share videos}}}\\
        \midrule
        
        \multirow{6}{*}{\makecell{Face of \\ test-taker}} & Blurring and Deepfake (O)  & $\chi^2$(1)=210.76, p<0.001 & $\chi^2$(1)=30.83, p<0.001  & $\chi^2$(1)=44.47, p<0.001  & $\chi^2$(1)=243.76, p<0.001& $\chi^2$(1)=158.20, p<0.001\\
        
        & Blurring and Deepfake (C)  & $\chi^2$(1)=215.98, p<0.001 & $\chi^2$(1)=29.32, p<0.001  & $\chi^2$(1)=52.28, p<0.001  & $\chi^2$(1)=255.01, p<0.001 & $\chi^2$(1)=320.56, p<0.001\\
        
        & Blurring and 3D avatar & $\chi^2$(1)=1.15, p=0.28 & $\chi^2$(1)=541.22, p<0.001 & $\chi^2$(1)=278.35, p<0.001  & $\chi^2$(1)=635.04, p<0.001 & $\chi^2$(1)=44.54, p<0.001\\
        
        & Deepfake (O) and Deepfake (C) & $\chi^2$(1)=0.03, p=0.86 & $\chi^2$(1)=0.02, p=0.89 & $\chi^2$(1)=0.32, p=0.57  & $\chi^2$(1)=0.13, p=0.72 & $\chi^2$(1)=28.37, p<0.001\\
        
        & Deepfake (O) and 3D avatar  & $\chi^2$(1)=180.80, p<0.001 & $\chi^2$(1)=313.71, p<0.001& $\chi^2$(1)=100.30, p<0.001  & $\chi^2$(1)=91.91, p<0.001 & $\chi^2$(1)=34.86, p<0.001\\
        
        & Deepfake (C) and 3D avatar  & $\chi^2$(1)=185.63, p<0.001 & $\chi^2$(1)=318.61, p<0.001 & $\chi^2$(1)=89.37, p<0.001  & $\chi^2$(1)=85.21, p<0.001 & $\chi^2$(1)=126.12, p<0.001\\

        \midrule
        
        {\makecell{Body of \\ test-taker}} & Blurring and Deepfake   & $\chi^2$(1)=17.94, p<0.001 & $\chi^2$(1)=66.30, p<0.001 & $\chi^2$(1)=62.12, p<0.001  & $\chi^2$(1)=150.45, p<0.001 & $\chi^2$(1)=18.18, p<0.001\\      
        
        \midrule
        
        {\makecell{Background \\ of test-taker}} & Blurring and Deepfake   & $\chi^2$(1)=65.39, p<0.001 & $\chi^2$(1)=467.86, p<0.001& $\chi^2$(1)=309.18, p<0.001  & $\chi^2$(1)=448.93, p<0.001 & $\chi^2$(1)=1.03, p=0.31\\
         
        \midrule
        
        \multirow{3}{*}{\makecell{People in \\ background}} & Blurring and Silhouette  & $\chi^2$(1)=116.75, p<0.001 & $\chi^2$(1)=218.63, p<0.001 & $\chi^2$(1)=136.85, p<0.001 & $\chi^2$(1)=133.34, p<0.001 & $\chi^2$(1)=10.09, p<0.005\\
        
        & Blurring and 3D avatar  & $\chi^2$(1)=29.50, p<0.001 & $\chi^2$(1)=25.06, p=0.001  & $\chi^2$(1)=7.69, p<0.01  & $\chi^2$(1)=54.29, p<0.001 & $\chi^2$(1)=110.94, p<0.001\\
        
        & Silhouette and 3D avatar   & $\chi^2$(1)=263.61, p<0.001 & $\chi^2$(1)=95.65, p<0.001 & $\chi^2$(1)=79.65, p<0.001  & $\chi^2$(1)=17.47, p<0.001 & $\chi^2$(1)=54.12, p<0.001\\
 
    \bottomrule
    \end{tabular*}
    \end{minipage}}
     \end{minipage}
\end{table}

\label{sec:appendix}

\end{document}